\documentclass[%
 reprint,twocolumn,
 groupedaddress,
 superscriptaddress,
 amsmath,amssymb,
 aps,
 prx,
]{revtex4-1}

\usepackage{graphicx}
\usepackage{mathrsfs}
\usepackage{bm}
\usepackage{bbold}
\usepackage[usenames,dvipsnames]{color} 
\usepackage[colorlinks=true , citecolor=blue,urlcolor=blue]{hyperref}
\usepackage{CJKutf8}
\newcommand{\lr}[1]{\left( #1\right)}

\newcommand{\comment}[1]{}
\newcommand{\mO}{\mathcal{O}}

\begin{document}

\begin{CJK}{UTF8}{gbsn}
\title{
Prethermal quasiconserved observables in Floquet quantum systems
}

\author{Chao Yin}\email[]{yinchao1998@pku.edu.cn}
\affiliation{
Research Laboratory of Electronics, Massachusetts Institute of Technology, Cambridge, Massachusetts 02139, USA
}
\author{Pai Peng (彭湃)}\thanks{C.Y. and P.P. contributed equally to this work.}
\affiliation{Department of Electrical Engineering and Computer Science, Massachusetts Institute of Technology, Cambridge, MA 02139}
\author{Xiaoyang Huang}
\affiliation{
Research Laboratory of Electronics, Massachusetts Institute of Technology, Cambridge, Massachusetts 02139, USA
}
\author{Chandrasekhar Ramanathan}
\affiliation{Department of Physics and Astronomy, Dartmouth College, Hanover, NH 03755, USA}
\author{Paola Cappellaro}\email[]{pcappell@mit.edu}
\affiliation{Department of Nuclear Science and Engineering, Massachusetts Institute of Technology, Cambridge, MA 02139}
\affiliation{
Research Laboratory of Electronics, Massachusetts Institute of Technology, Cambridge, Massachusetts 02139, USA
}

\date{\today}

\begin{abstract}
Prethermalization, by introducing emergent quasiconserved observables,  plays a crucial role in protecting periodically driven (Floquet) many-body phases over exponentially long time, while the ultimate fate of such quasiconserved operators can signal  thermalization to infinite temperature. To elucidate the properties of  prethermal quasiconservation in many-body Floquet systems, here  we systematically analyze infinite temperature correlations  between observables. We numerically show that the late-time behavior of the autocorrelations unambiguously distinguishes  quasiconserved observables from non-conserved ones, allowing to single out a set of linearly-independent  quasiconserved observables. By investigating two Floquet spin models,  we identify two different mechanism underlying the quasiconservation law.
First, we numerically verify energy quasiconservation when the driving frequency is large, so that the system dynamics is approximately described by a static prethermal Hamiltonian. 
More interestingly, under moderate driving frequency, another quasiconserved observable can still persist if the Floquet driving contains a large global rotation. We show theoretically how to calculate this conserved observable and provide numerical verification. Having systematically identified all quasiconserved observables, we can finally investigate their behavior  in the infinite-time limit and thermodynamic limit, using autocorrelations obtained from both numerical simulation and experiments in solid state nuclear magnetic resonance systems.
\end{abstract}
\maketitle
%
\end{CJK}

\section{Introduction}
Controlling quantum systems using a periodic (Floquet) drive has emerged as a powerful tool in the field of condensed matter physics and quantum information science. It has been used to realize Hamiltonians that are not accessible in a static system, such as modifying the tunneling and coupling rates~\cite{Eckardt05,Tsuji11,Mentink15,Kitamura16,Mikhaylovskiy15,Gorg18}, inducing non-trivial topological structures~\cite{Lindner11,Wang13s,Oka09,Gu11,Grushin14,FoaTorres14,Rudner13,Jiang11l,Kundu13,Kitagawa10,Else17b}, creating synthetic gauge fields~\cite{Goldman14,Bukov15,Bukov16,Struck12,Aidelsburger13} and spin-orbit couplings~\cite{Struck14}. On a quantum computer, Floquet engineering also enables universal quantum simulation via Trotter-Suzuki scheme~\cite{trotter59,Lloyd96,liu19x,Jotzu14,Aidelsburger15,Kokail19,Childs2018}. Floquet systems also possess interesting dynamical phenomena ranging from discrete time crystalline phase~\cite{Choi17n,Zhang17n,Moessner17,Luitz19,Machado19x} to dynamical localization~\cite{Dunlap86,Fishman82}, dynamical phase transitions~\cite{Bastidas12,Bastidas12a} and coherent destruction of tunneling~\cite{Grossmann91,Grossmann92,Grifoni98}.  

While the connection to an effective time-independent Hamiltonian is appealing, the active drive leads to energy absorption by the Floquet many-body system, which is then expected to heat up to infinite temperature. The heating is detrimental to any quantum application, as no local quantum information is retained and all interesting phenomena mentioned above disappear~\cite{Lazarides14,DAlessio14,Kim14}.
It has been shown theoretically  ~\cite{Abanin17,Abanin15,Kuwahara16,Abanin17b,Else17} and experimentally~\cite{Peng2019,Antonio2020} that even when the system heats up,  the thermalization time can be exponentially long in  the drive parameters (typically the frequency of a rapid drive). Then, a long-lived \textit{prethermal} quasi-equilibrium is established, that allows exploiting the engineered Floquet Hamiltonian for quantum simulation~\cite{Heyl19,DAlessio13,Sieberer19}. The emergent symmetries and conserved observables in the prethermal state distinguish it from the fully thermalized state, and underpin the existence of novel Floquet phases~\cite{Else17,Luitz19,Machado19x}.
Even more surprisingly, some numerical studies have shown that the emergent conserved observables might not display thermalizing behavior  even in the infinite-time limit~\cite{Heyl19,Sieberer19,Prosen99,DAlessio13}. Many-body localization~\cite{Abanin16,Lazarides15,Ponte15,Zhang17n,Zhang16b,Po16,Bordia17,Khemani16}, dynamic localization~\cite{Heyl19,Sieberer19,Ji18}, and some fine-tuned driving protocols~\cite{Prosen98,Prosen99,DAlessio13} provide a way to escape the thermalization fate, which could also be absent in finite-size systems.
Indeed, distinguishing the long-lived prethermal state from an eventual thermal state is challenging. Numerical studies are bound to finite-size (and often small) systems, while experiments can only probe finite times, before the external environment induces thermal relaxation. 

Here we tackle this problem by a numerical and experimental study of two Floquet  models in spin chains, namely the kicked dipolar model (KDM) and the alternating dipolar model (ADM). While most studies on spin chain dynamics have focused on evolution of pure states, 
here we propose to study Floquet prethermalization using infinite temperature correlations.
This metric provides information about quasisconserved observables across the whole spectrum and serves as a direct measurable quantity in nuclear magnetic resonance (NMR) experiments. 
In Sec.~\ref{sec:conserv} we show that the existence of long-lived quasiconserved observables can be unambiguously identified using late-time behavior of the correlations, based on which we provide a method to systematically search for all linearly-independent local quasiconserved quantities. 
Then we provide both numerical and analytical tools to investigate such prethermal conserved observables and their origins. We first show that the prethermal Hamiltonian $H_{pre}$ obtained from the Magnus expansion under rapid drive yields a quasiconserved observable in each model in Sec.~\ref{sec:Hpre}. We further show in Sec.~\ref{sec:Dpre} that when the driving Hamiltonian contains a large global rotation, the Floquet propagator can induce an additional conserved observable, as shown by going beyond the usual Magnus expansion.
With all the quasiconserved observables at hand, we investigate in Sec.~\ref{sec:limit} whether they exist in the thermodynamic limit and infinite-time limit, by looking at the dependence of autocorrelations on system size (numerically) and on time (experimentally). Both methods indicate quasiconserved observables vanish and the system thermalizes to infinite temperature.

\section{Quasiconserved observables}
\label{sec:conserv}
\subsection{Hamiltonians and Correlations}
\begin{figure}[b]
\centering
\includegraphics[width=0.98\linewidth]{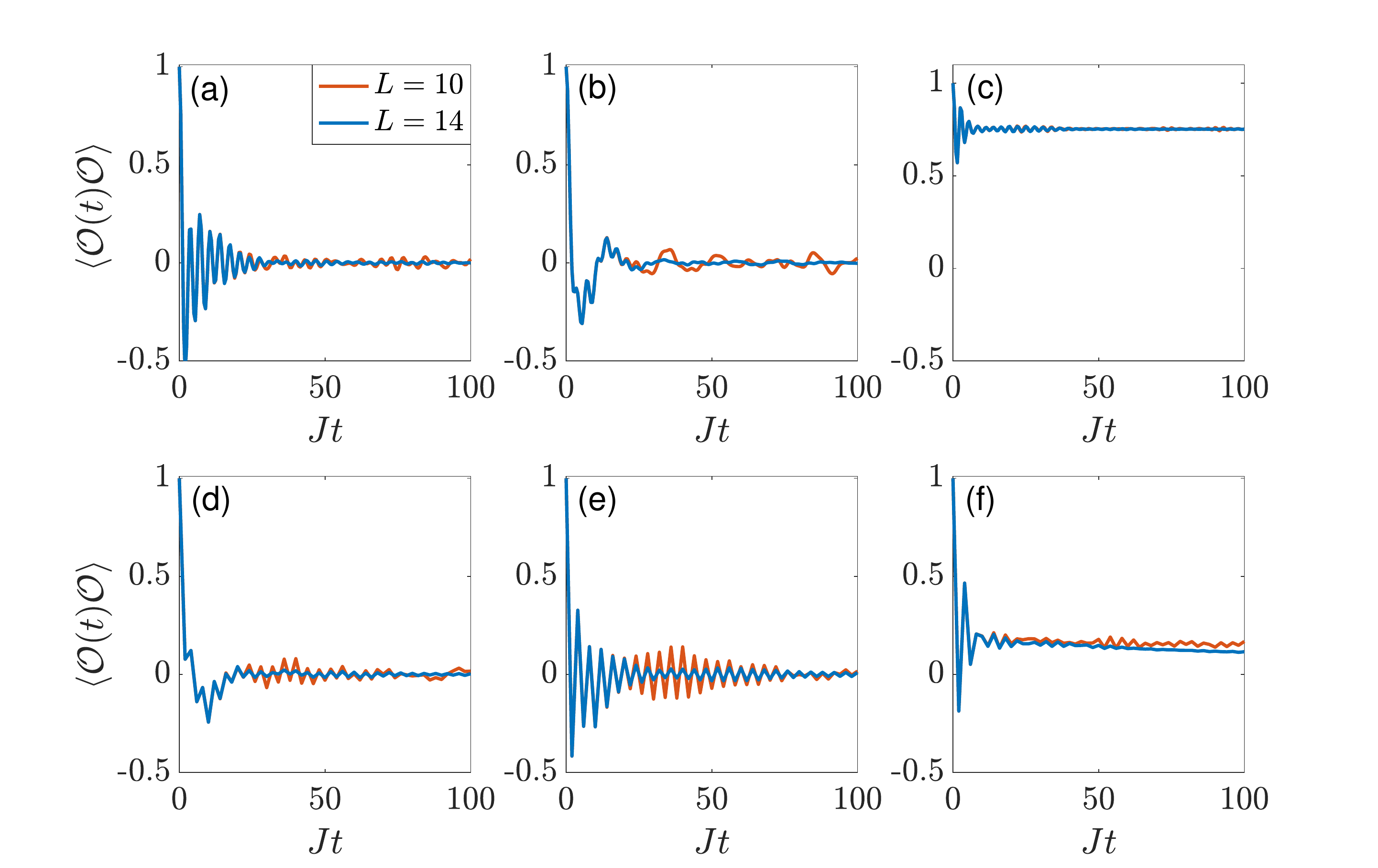}
\caption{\label{fig:dyn}
Typical dynamics of $\langle \mO(t){\mO}' \rangle$ in a Floquet spin chain. Here we choose KDM and $\mO=\mO'$. (a-c) $J\tau=0.5$, (d-f) $J\tau=2$. (a,d) $\mO=X$, (b,e) $\mO=Y$, (c,f) $\mO=Z$. Different colors correspond to different system size $L$, as shown in the legend.
}
\end{figure}

In this paper we use the Trotter-Suzuki scheme for the driving protocol, where the time-dependent Hamiltonian is piecewise constant in one driving period. However, our results are general for any form of periodic driving. The evolution of the system we study is given by the unitary propagator in one period $U_F=e^{-iH_2\tau}e^{-iH_1\tau}$, where in each period we consider the system to be under the Hamiltonian $H_1$ for a time $\tau$, and then under $H_2$ for another duration $\tau$.
Motivated by NMR experiments, we consider two models of an $L$-site spin-1/2 chain: the kicked dipolar model (KDM), where $H_1^{(K)}\!=\!JD_y$, $H_2^{(K)}\!=\!hZ$, and the alternating dipolar model (ADM), with $H_1^{(A)}\!=\!JD_y$ and $H_2^{(A)}\!=\!JD_x$.
Here $D_\alpha=\sum_{j<k} \frac{1}{2}\left(3S_\alpha^j S_\alpha^k - \vec{S}_j\cdot\vec{S}_k \right)/|j-k|^3$ is the dipolar interaction operator in an arbitrary direction set by $\alpha$ \((\alpha=x,y,z)\), where \(S_\alpha^j\)  are spin-1/2 operators of the \(j\)-th spin $(j=1,\cdots,L)$ and $\vec{S}_j=(S_x^j,S_y^j,S_z^j)^T$. As shown in Ref.~\cite{Machado19x}, the $1/r^3$ interaction is sufficiently short range in 1D to yield no qualitative difference with respect to the nearest-neighbor interaction, thus for simplicity in numerical and analytical studies we only keep  the nearest-neighbor interaction  unless explicitly mentioned. $Z=\sum_j S^j_z$ is the
collective magnetization operator along z-axis, and below we will also use $X=\sum_j S^j_x, Y=\sum_j S^j_y$. $J$ and $h$ are the  strength of the dipolar interaction and the collective z-field respectively, and we fix $h=J$ throughout the paper. In numerics we assume periodic boundary conditions.

To investigate quasi-conservation properties we use  infinite-temperature correlations as our metric, $\langle \mO(t){\mO}' \rangle_{\beta=0} \equiv \text{Tr}[ U_t\mO U_t^\dagger{\mO}' ] /\lr{\| \mO\|\|{\mO}'\| }$, where $U_t$ is the unitary evolution during time $t$, ${\mO}$ and ${\mO}'$ are observables, and the norm is defined as $\|\mO\|\equiv \sqrt{\text{Tr}\mO^2}$. In the following we drop the subscript $\beta=0$ for simplicity.

Figure~\ref{fig:dyn} shows numerical simulations of  some exemplary correlations, the magnetization along three axes $\mO={\mO}'=Z,X,Y$ in KDM (the qualitative behavior is general for other observables and models.) 
The autocorrelations of $X$ and $Y$ display  oscillations around $0$ and damping, which originate from the z-field and the dipolar interaction, respectively. 
Instead, $\langle Z(t)Z \rangle$ exhibits a more interesting behavior. For small $J\tau$, it quickly equilibrates at a nonzero value independent of $L$, and it remains constant afterwards. For relatively large $J\tau$, there is a slow decay of $\langle Z(t)Z \rangle$ toward a final value that decreases with increasing $L$. We thus expect the final value to be zero in the thermodynamic limit, corresponding to an infinite-temperature final state. 
Indeed, the observable $Z$ displays the defining characteristics of what we deem a quasiconserved observable in the prethermal regime: the autocorrelation of a quasiconserved observable is nonzero in the prethermal regime, but goes to zero in the fully thermalized state. 
In simulations,  autocorrelations of quasiconserved observables still have nonzero value at infinite time due to the small system size (e.g. $\langle Z(t)Z\rangle$ in Fig.~\ref{fig:dyn}), while for non-conserved observables autocorrelations are zero (e.g. $\langle X(t)X\rangle$ in Fig.~\ref{fig:dyn}). These distince behaviors serve as a direct metric to identify quasiconserved observables. As any observable that overlaps with a quasiconserved observable would have non-zero infinite-time autocorrelation, we want to find a linearly independent, orthogonal set of  \textit{eigen-}quasiconserved observables.

\subsection{Eigen-quasiconserved Observables}

We design a systematic procedure to search for the set of eigen-quasiconserved observables, $\{\mathcal E_\mu\}$ starting from the infinite-time correlations $\langle \mO(\infty)\mO' \rangle\equiv\lim_{T\to\infty}(1/T)\int_0^T\langle \mO(t)\mO' \rangle dt$. 
We note that eigenvectors $\{E_\mu\}$  of the Floquet (super)propagator $\hat U_{F}$ form an orthogonal vector basis for the space of operators (here $\hat U[\mathcal O]=U\mO U^\dagger$.),  $|\langle  E_j(\infty)E_k\rangle| \propto \delta_{jk}$, that we can call ``eigen-observables''. 

However, this operator basis is in general highly non-local, and thus not practical.
We then want to find a small, local set of  observables that approximate the exact eigen-observables, and have non-zero eigenvalues, that is, are quasiconserved. 
We start from a basis set $\{\mO_{(\alpha)}\}$ of Hermitian observables  that are translationally invariant sums of local operators:
\begin{equation}
    \mO_{(\alpha)} = \sum_j S^j_{\alpha_1}S^{j+1}_{\alpha_2}\cdots S^{j+r-1}_{\alpha_r}.
\end{equation}
Here $(\alpha)\equiv(\alpha_1, \cdots, \alpha_r)$ with  $\alpha_{k}\in\{x,y,z,0\}$, where $S_0^j$ denotes the identity matrix operating on the $j$-th spin. By imposing $\alpha_1,\alpha_r\neq0$, we say $\mO_{(\alpha)}$ is of range $r$:
each term in $\mathcal O_{(\alpha)}$ acts non-trivially on at most $r$ neighboring spins. 
Since the number of operators is exponentially large in system size, we restrict our search to the operator subspace spanned by $\mO_{(\alpha)}$ whose range $r\le r_c$, which are local and thus experimentally relevant. 
Starting from an orthonormal operator basis $\{{\mO}_\mu\}$ of this subspace (with  $\langle{\mO}_\mu{\mO}_\nu\rangle=\delta_{\mu\nu}$)
we construct a matrix from all pair correlations,  $\Lambda_{\mu\nu}=\langle{\mO}_\mu(\infty){\mO}_\nu\rangle$.
The matrix $\Lambda$ is the projection of the infinite-time propagator $\hat U_F(t\to\infty)$ onto the $r_c$-local subspace. 
The diagonalization of $\Lambda$ yields  the local eigen-observables $\mathcal E_k$, and eigenvalues $\lambda_k$, satisfying $\langle\mathcal{E}_k(\infty)\mathcal{E}_l\rangle=\lambda_k\delta_{kl}$. Note that since $\Lambda$ is not ensured to be unitary, its eigenvalues do not have unit amplitude, $\lambda_k\leq 1$. We note that the larger the $\lambda_k$, the better $\mathcal E_k$ approximates an exactly conserved observable. 
{The correlations  $\langle \mO(\infty){\mO'} \rangle$ between any two observables whose locality is bounded by $r_c$ can be directly derived by  decomposing the observables onto the $\mathcal E_\mu$ basis
\begin{equation}\label{eq:decomp}
    \langle \mO(\infty){\mO}' \rangle = \sum_\mu \lambda_\mu \langle\mO \mathcal E_\mu\rangle \langle\mathcal E_\mu{\mO'}\rangle.
\end{equation} }

\begin{figure}[!htp]
\centering
\includegraphics[width=0.48\linewidth]{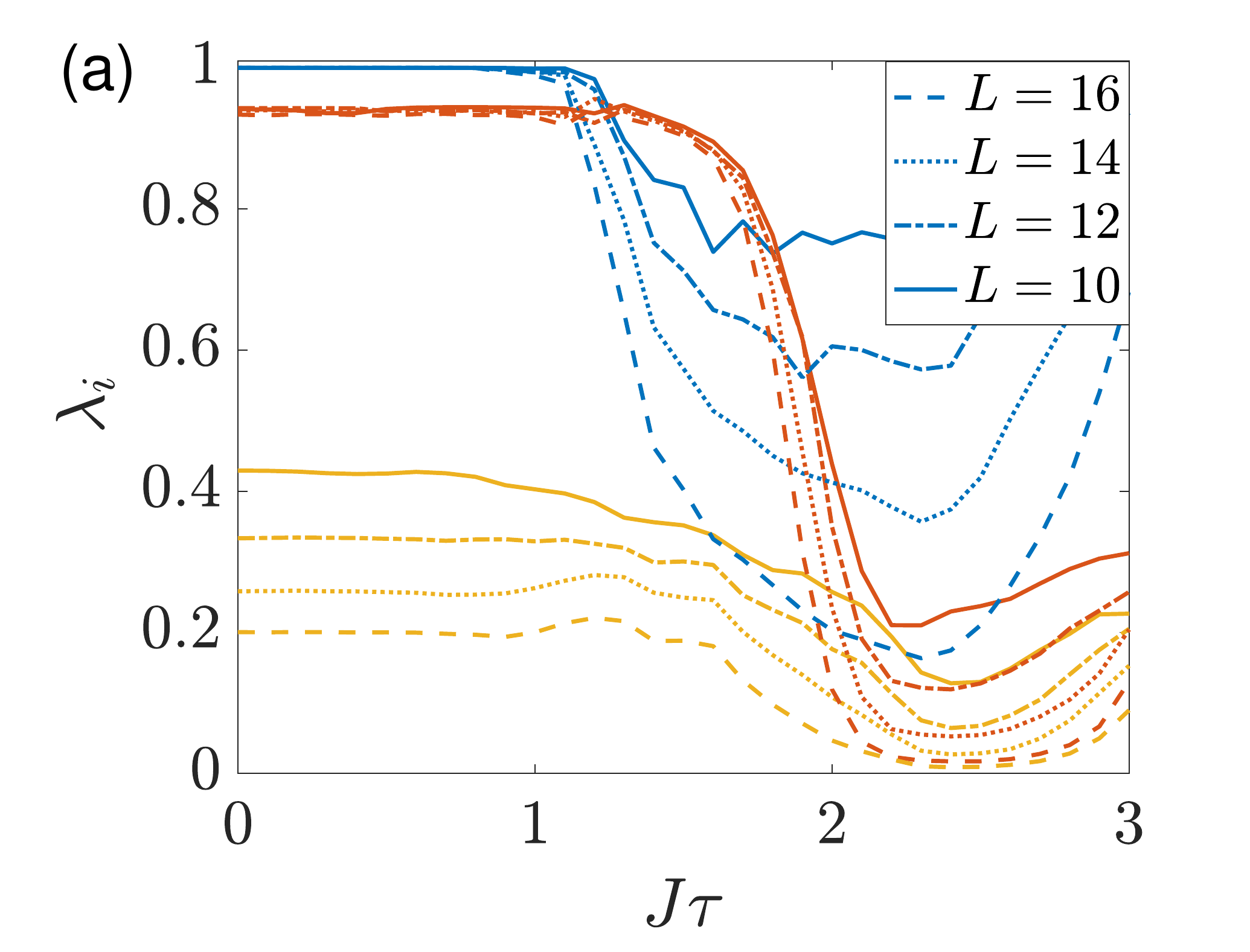}
\includegraphics[width=0.48\linewidth]{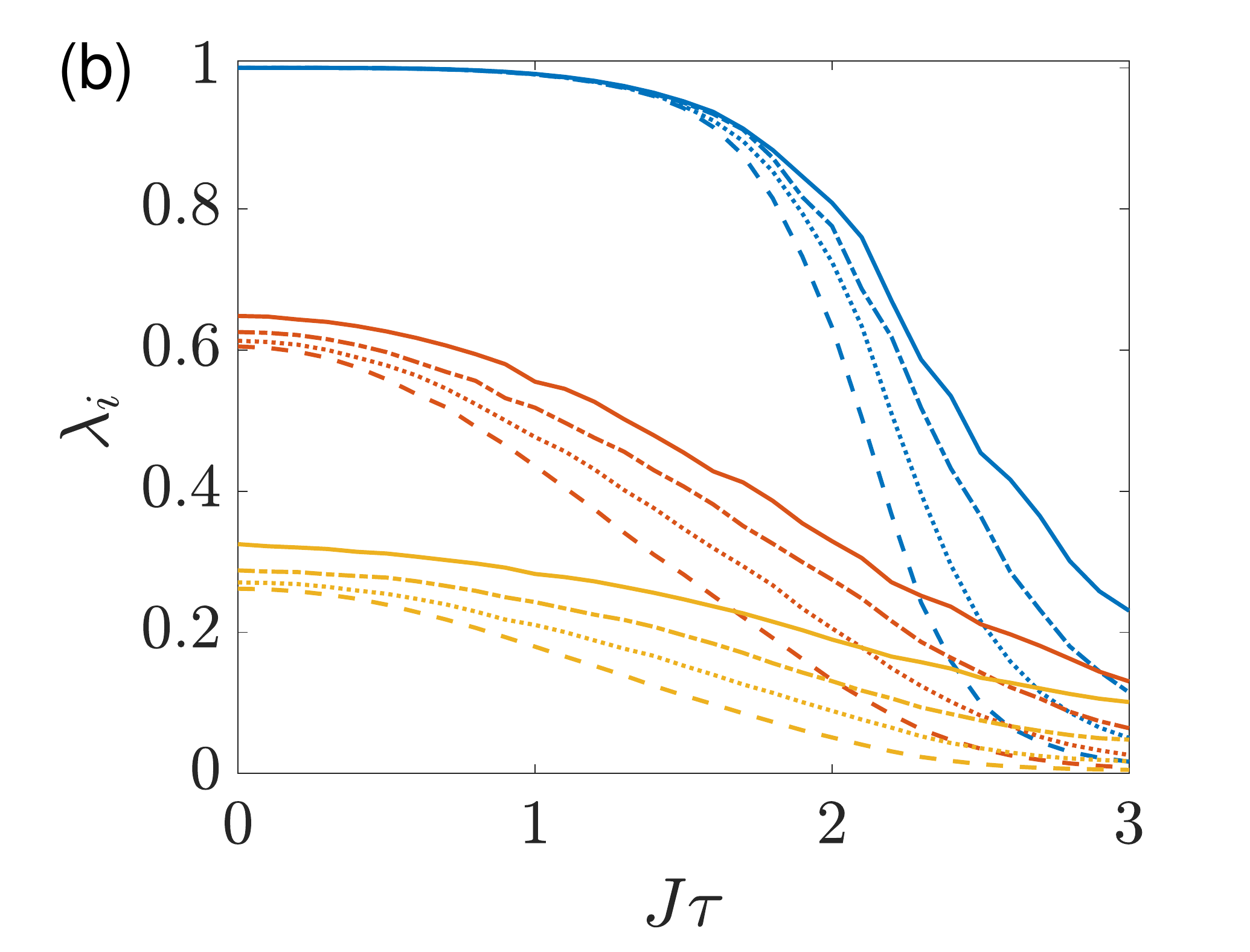}
\caption{\label{brute}
 By considering the matrix $\Lambda$ obtained for each $J\tau$ Trotter step, we calculate three largest eigenvalues as a function of $J\tau$ for KDM (a) and ADM (b). Curve color represents different eigenvalues and curve style represents different system sizes. From the eigenvalues and their dependence on system size, we see there are two eigen-quasiconserved observables in KDM while only one in ADM.
}
\end{figure}

We apply this systematic procedure to the two models under consideration. The infinite time limit $\mathcal{O}(\infty)$ is taken by considering the diagonal ensemble of  $\mathcal{O}$ (that is, keeping only the diagonal matrix elements of $\mathcal{O}$ in the Floquet energy eigenbasis), which gives the same result as averaging $\mathcal{O}$ over long time. The results for $r_c=3$ are shown in Fig.~\ref{brute}.
At large Trotter steps, $\tau$, most eigenvalues go to zero. The upward trends of the eigenvalues  when $J\tau=h\tau\to\pi$ (most pronounced for the largest  eigenvalue) is due to the fact that  $[e^{-iH_1^{(K)}\tau},e^{-iH_2^{(K)}\tau}]=0$ at $J\tau=h\tau\to\pi$, making the system  equivalent to a time-independent system.
Even for small Trotter steps, most eigenvalues are already small, and decrease when increasing system size. 
However, a few eigenvalues are large,  and show little dependence on system size.
This last group comprises the eigenvalues associated with  the eigen-quasiconserved observables that govern the nontrivial dynamics at long times.

Based on these results, we find that there are two eigen-quasiconserved observables for KDM, $\mathcal E^{(K)}_1,\mathcal E^{(K)}_2$,  and one for ADM, $\mathcal E^{(A)}_1$. 
In both models, $\mathcal E_1$ is close to their average Hamiltonian $\overline H=H_1+H_2$ (blue curves in Fig.~\ref{brute}), while $\mathcal E^{(K)}_2$ for KDM is close to $D_z$ [red curves in Fig.~\ref{brute}(a)]. We can thus more carefully analyze these quasiconserved observables and describe analytically their origin in the limit of small $\tau$ in the next section. 
Even so, we remark that there is an interesting regime at intermediate $\tau$
, where $\mathcal E^{(K)}_1,\mathcal E^{(K)}_2$ are well conserved, since $\lambda^{(K)}_1,\lambda^{(K)}_2$ are  still large, but they deviate from their static ($\tau\to0$) counterparts. This indicates that the quasiconserved observables truly arise from the Floquet dynamics, and are not simply a remnant of the approximated, static Hamiltonian.

\section{Analytical Derivation of Conserved Observables}      
\subsection{Prethermal Hamiltonian}\label{sec:Hpre}

\begin{figure*}
\centering
\includegraphics[width=0.32\textwidth]{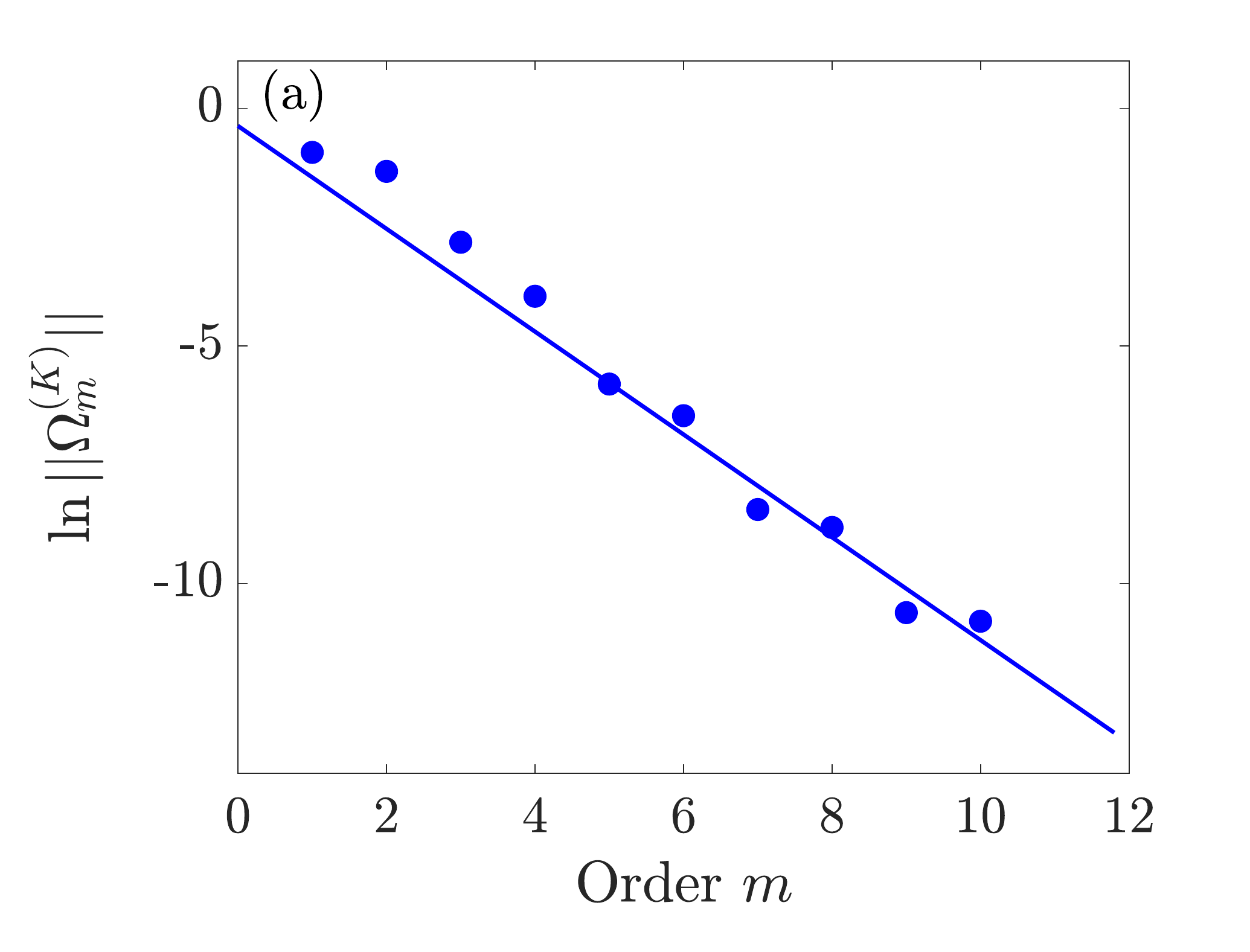}
\includegraphics[width=0.32\textwidth]{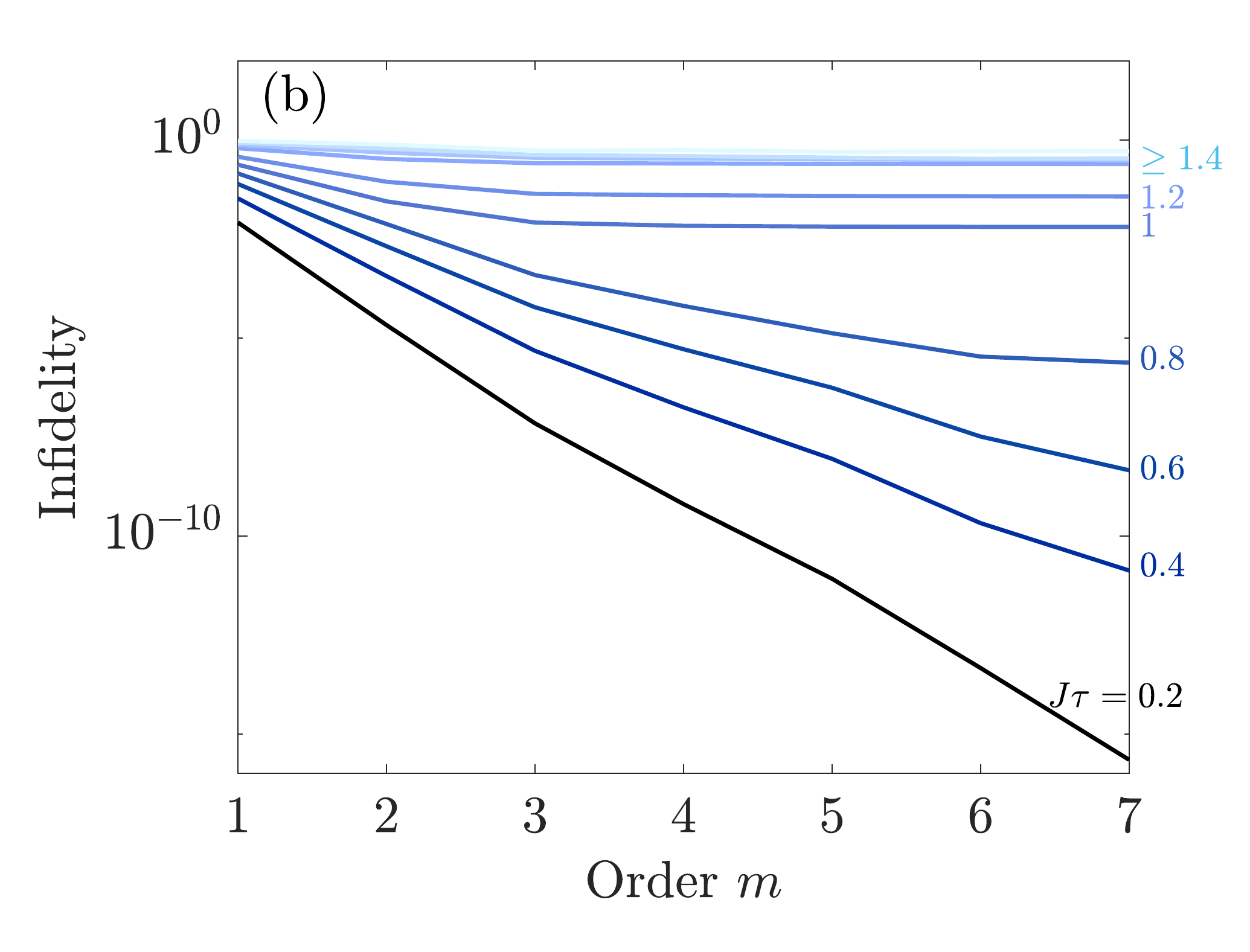}
\includegraphics[width=0.32\textwidth]{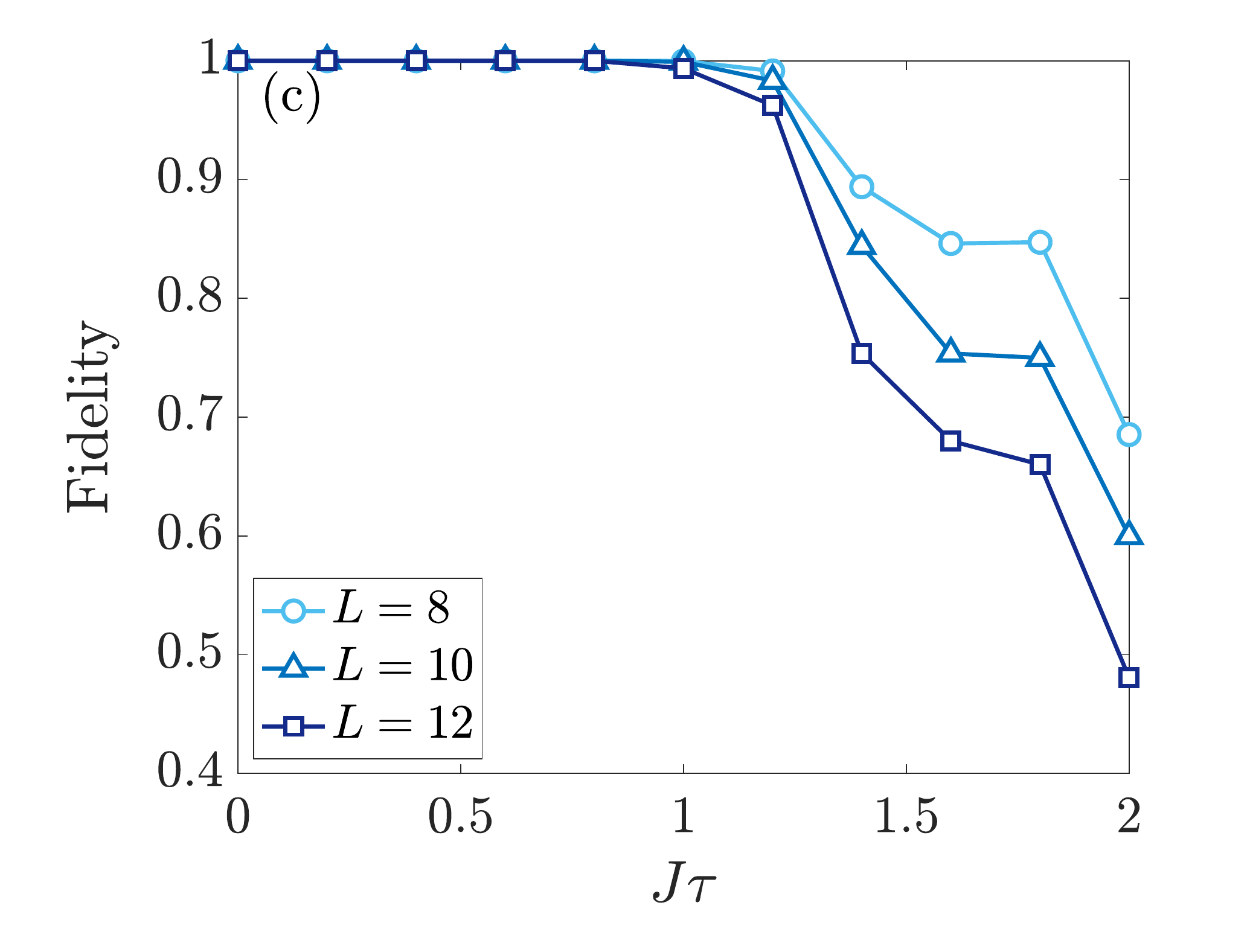}
\includegraphics[width=0.32\textwidth]{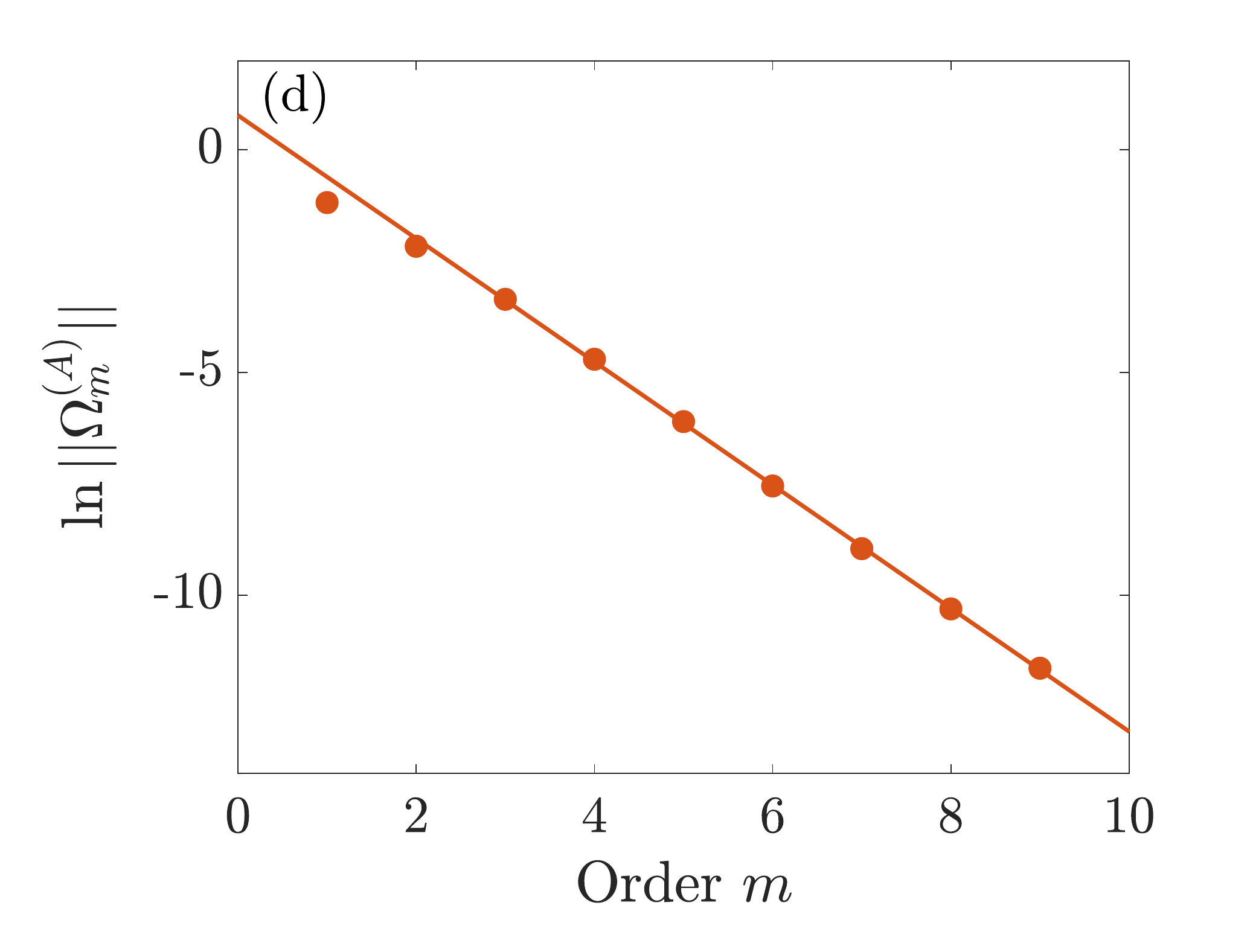}
\includegraphics[width=0.32\textwidth]{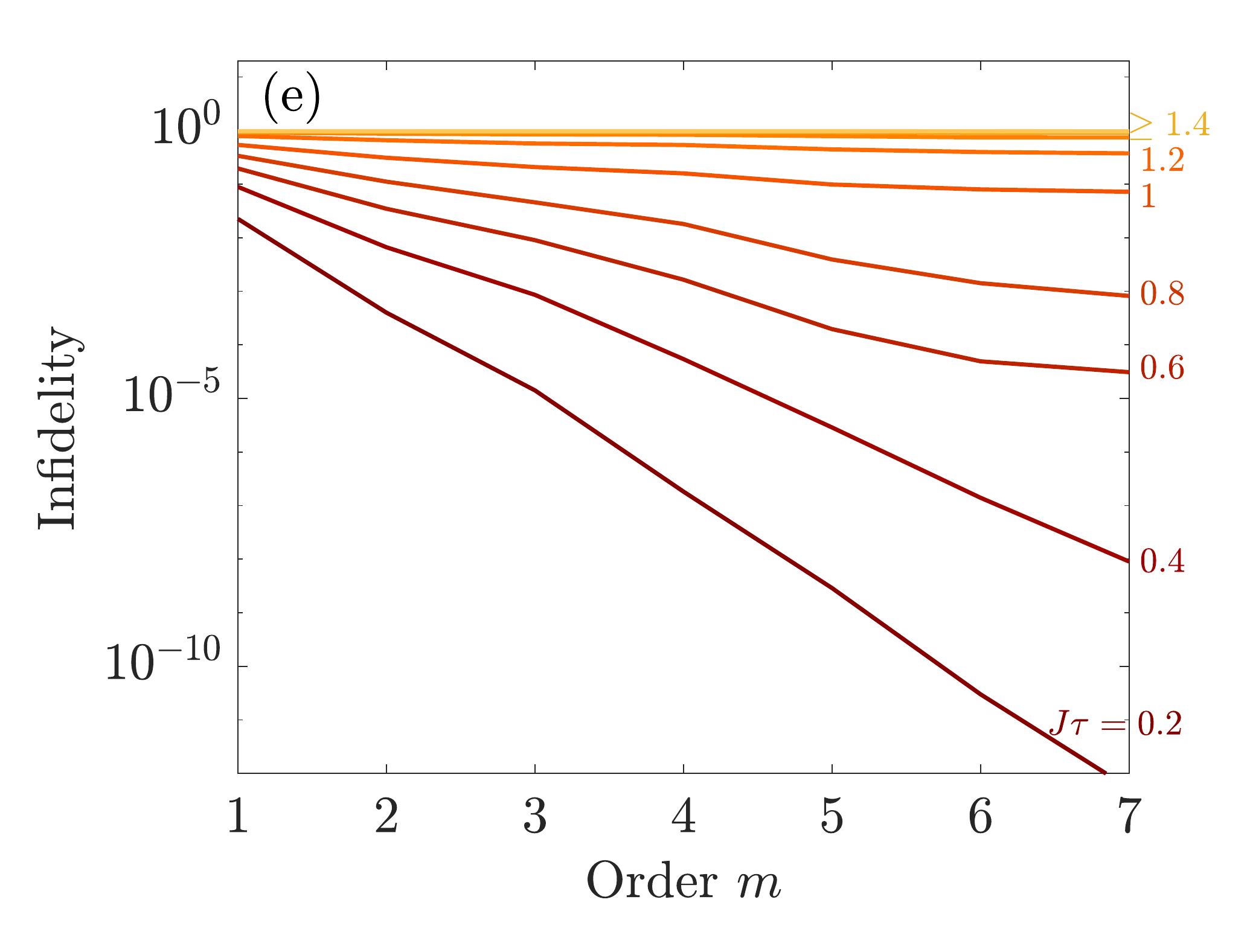}
\includegraphics[width=0.32\textwidth]{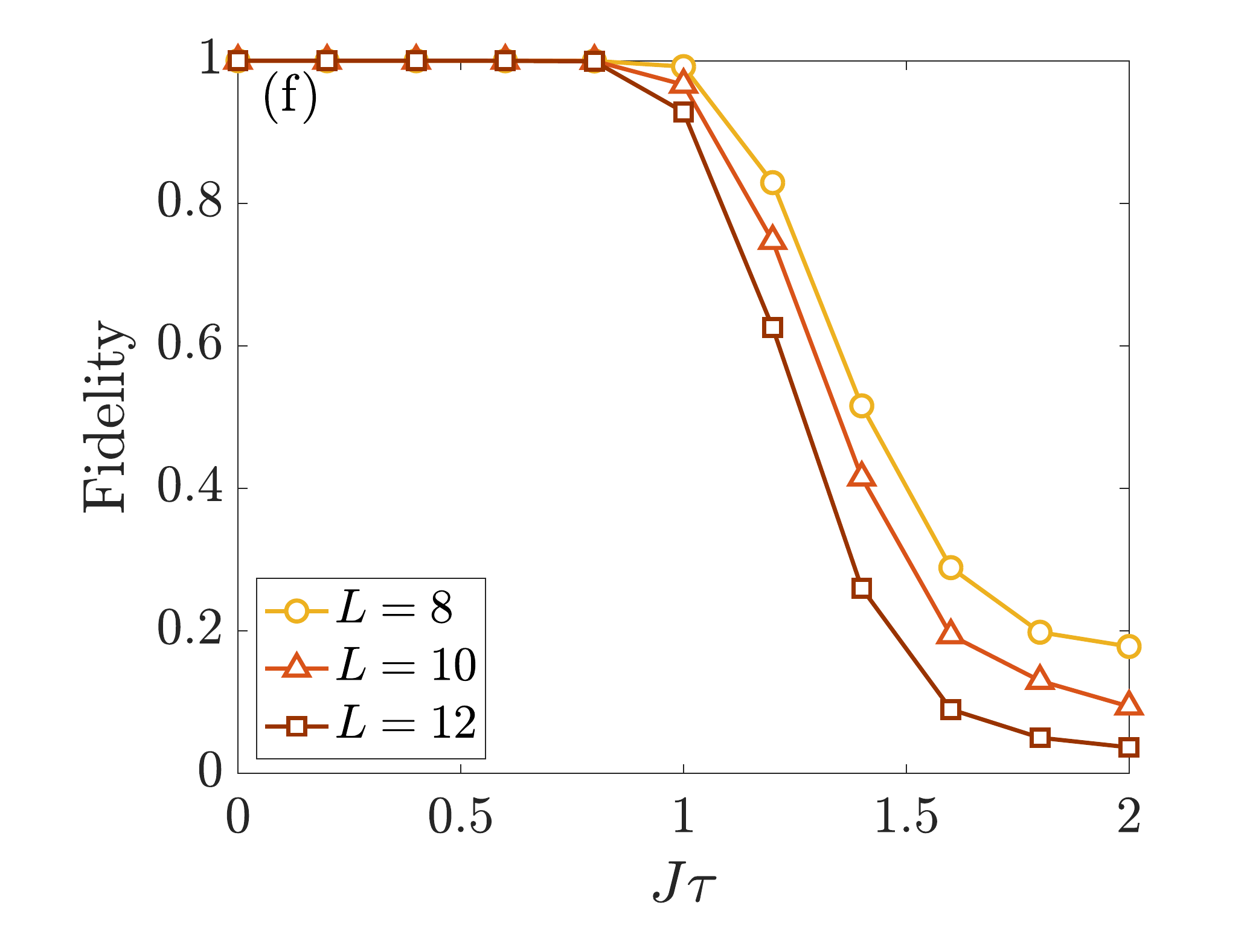}
\caption{\label{Hid_theory}
(a) to (c) show the Magnus expansion Eq.~\ref{eq:Hpre} of KDM, and (d) to (f) show that of ADM. (a) (d) Circles show the norm of $\Omega_m$ (normalized by $L2^L$).
Solid line represents the linear fit. (b) (e) infidelity $1-\langle H_{pre}(\infty)H_{pre}\rangle$ of infinite-time averaged $H_{pre}$ evaluated up to $m^\text{th}$ order. Different curves stand for $J\tau$ from 0.2 to 2 with a step of 0.2. Darker color represents  smaller $J\tau$. $L=12$ is used.
(c) (f) infinite-time autocorrelation of $H_{pre}$ as a function of $J\tau$ for different system sizes. Order $m=7$.
}
\end{figure*}

It is intuitive to expect that a quasiconserved observable might emerge from energy conservation. Indeed, one can always regard the Floquet evolution as arising from an effective static Hamiltonian by setting $U_F=e^{-i\tau H_F}$ for some Hermitian operator $H_F$.
However, in general  this Hamiltonian is highly non-local and thus it is not associated to a local quasi-conserved observable. 
Still, when the driving frequency is large compared to local energy scales (here $J,h$), the stroboscopic dynamics is given by a time-independent local prethermal Hamiltonian $H_{pre}$ plus a small correction $\delta H(t)$ ~\cite{Kuwahara16, Abanin17}, which may be nonlocal. 
It is this prethermal Hamiltonian $H_{pre}$ that can be associated with a local quasiconserved observable. 
$H_{pre}$ can be obtained from the Floquet-Magnus expansion~\cite{Magnus54, Blanes09} truncated at an optimal order $m^*$:
\begin{equation}\label{eq:Hpre}
H_{pre}=\sum_{m=0}^{m^*}\tau^m \Omega_m,
\end{equation}
where the zeroth order term is the average Hamiltonian $\Omega_0\!=\!\overline H\!=\!1/\tau\int_{0}^\tau\! H(t)dt$ and higher order terms $\Omega_m$ involve $m$ nested commutators. 
Then, for spin chains with {nearest-neighbor} couplings the range of $\Omega_m$ grows linearly with $m$.

The truncation $m^*$ is crucial not only to keep the prethermal Hamiltonian  local, but also because the series in Eq.~\ref{eq:Hpre} diverges for a generic many-body system~\cite{Kuwahara16}. 
The time-dependent correction $\delta H$ is however exponentially small in $1/J\tau$, leading to an exponentially long time $t_{pre}$ for the system to heat up. Thus, for $t<t_{pre}$, the system effectively prethermalizes to the state $e^{-\beta H_{pre}}$ where $\beta$ is determined by the initial state energy, making $H_{pre}$ an eigen-quasiconserved observable. 
Although one should investigate the prethermalization process by studying the dynamics of an infinitely large system at long times approaching infinity, numerically we can  only tackle small system sizes, so we take a different approach -- we set the time to infinity, and study how the observable correlations change when increasing system size.
The validity of this approach relies on the fact that for a system size $L<m^*$ the term $\delta H$ does not appear in the expansion, making $\mO_1= H_{pre}$ exactly conserved even at infinite time for sufficiently small $\tau$. From a physics point of view, this means that the energy  $2\pi\hbar/\tau$ is larger than the many-body bandwidth ($\sim JL$), and thus the system cannot absorb energy from the drive if it is faster than $1/JL$.
Since the zeroth order term of $H_{pre}$ is $\overline H$, the autocorrelation of $H_{pre}$ provides a bound for that of $\overline H$, leading to bounded Trotter error in the Trotter-Suzuki scheme \cite{Heyl19}. 

As further verification, we calculate numerically the  Floquet Magnus expansion, Eq.~(\ref{eq:Hpre}), up to $m=10$ and evaluate not only the convergence of the expansion, but also  operator conservation.
For the first metric, we plot $\|\Omega_m\|$ in Fig.~\ref{Hid_theory}(a) and (d) for the two models studied.
We find that, up to the computationally accessible order, the norm of $\Omega_m$  decays exponentially, indicating that $H_{pre}$ converges when $\tau$ is small.
From the slopes in Fig.~\ref{Hid_theory}(a) and (d), we get radii of convergence $J\tau\approx 3$ for both models.
Still, the expansion convergence does not guarantee the resulting $H_{pre}$ is a quasiconserved observable.
In Fig.~\ref{Hid_theory}(b) and (e), we compute the long-time infidelity ($1-\langle H_{pre}(\infty) H_{pre} \rangle$)
by truncating the expansion in Eq.~(\ref{eq:Hpre}) at increasing orders.
When $J\tau$ is small, the autocorrelation exponentially approaches 1 with increasing order, suggesting that the optimal truncation order $m^*$ should be larger than our largest accessible order here, or even absent in the system size we study. 
Instead, for larger $J\tau$, the correlation stops  converging at some order; for even larger $J\tau$ ($J\tau=1$ for example) the correlation  is almost zero for all orders. Therefore, even within the radius of convergence $J\tau\approx3$,  $H_{pre}$ from Eq.~\ref{eq:Hpre} may fail to be quasiconserved.
We plot the infinite-time correlation $\langle H_{pre}(\infty) H_{pre} \rangle$  versus $J\tau$ in Fig.~\ref{Hid_theory}(c) and (f) and show how it changes with system size (here $H_{pre}$ is evaluated to $7^\mathrm{th}$ order). The drop of $\langle H_{pre}(\infty) H_{pre} \rangle$ with increasing system size is evident for $J\tau\gtrsim1.2$ in both models, suggesting that for the system size we explore the effective Hamiltonian picture fails in the above parameter space. Note that in the $L\to\infty$ limit the correlations are expected to be zero for any $\tau>0$ as will be discussed in Sec.~\ref{sec:limit}.

\subsection{Emergent dipolar order}\label{sec:Dpre}

\begin{figure*}
\centering
\includegraphics[width=0.32\textwidth]{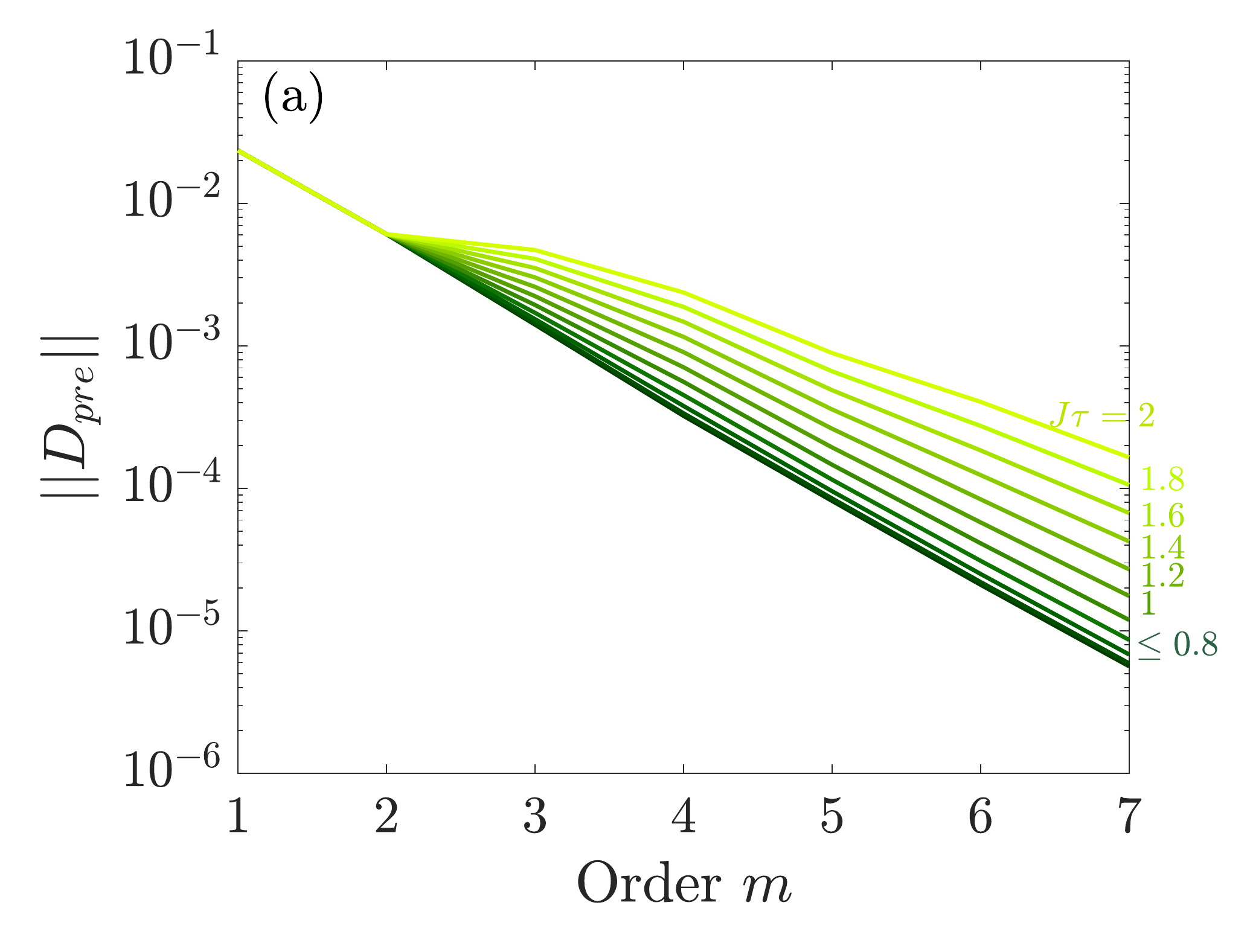}
\includegraphics[width=0.32\textwidth]{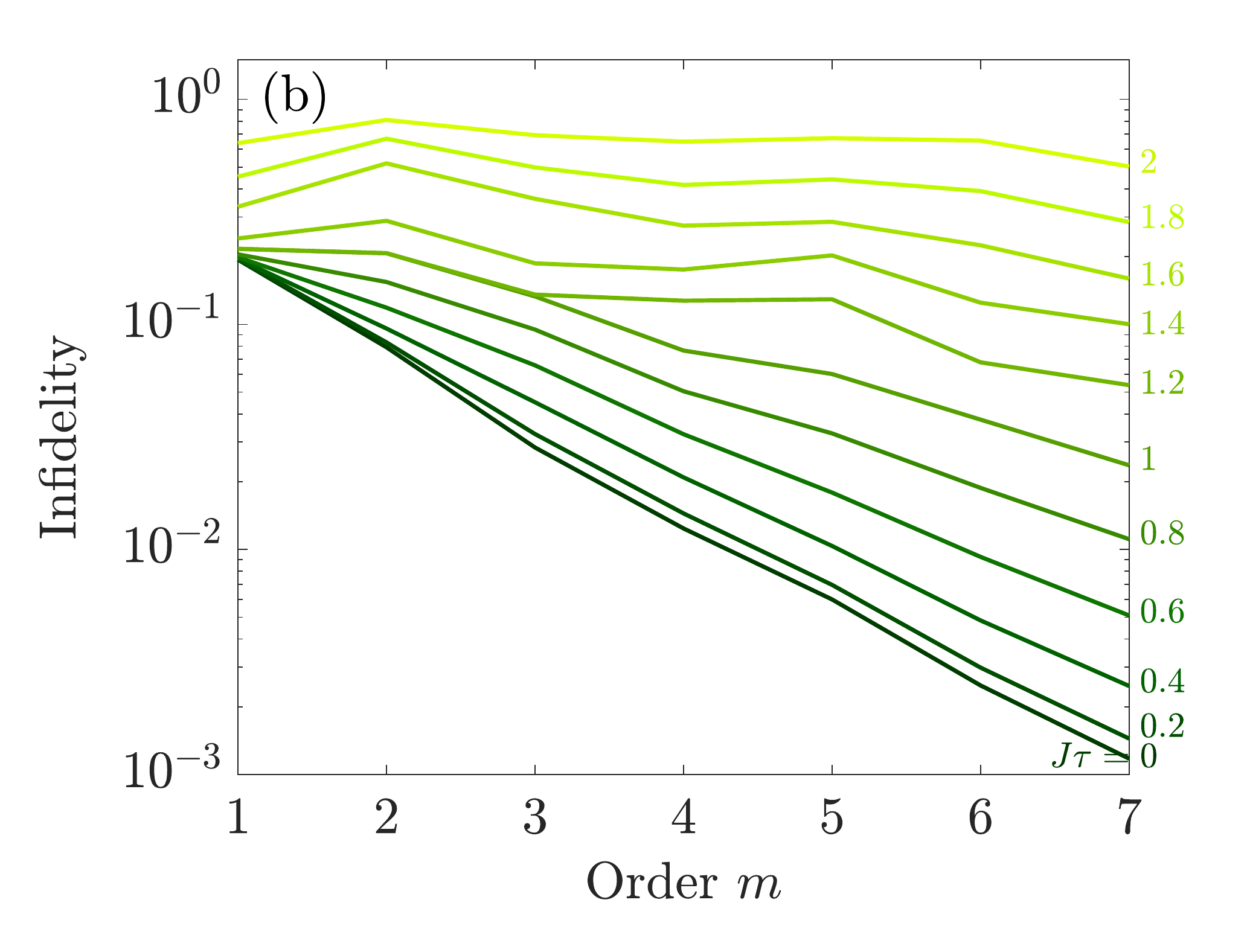}
\includegraphics[width=0.32\textwidth]{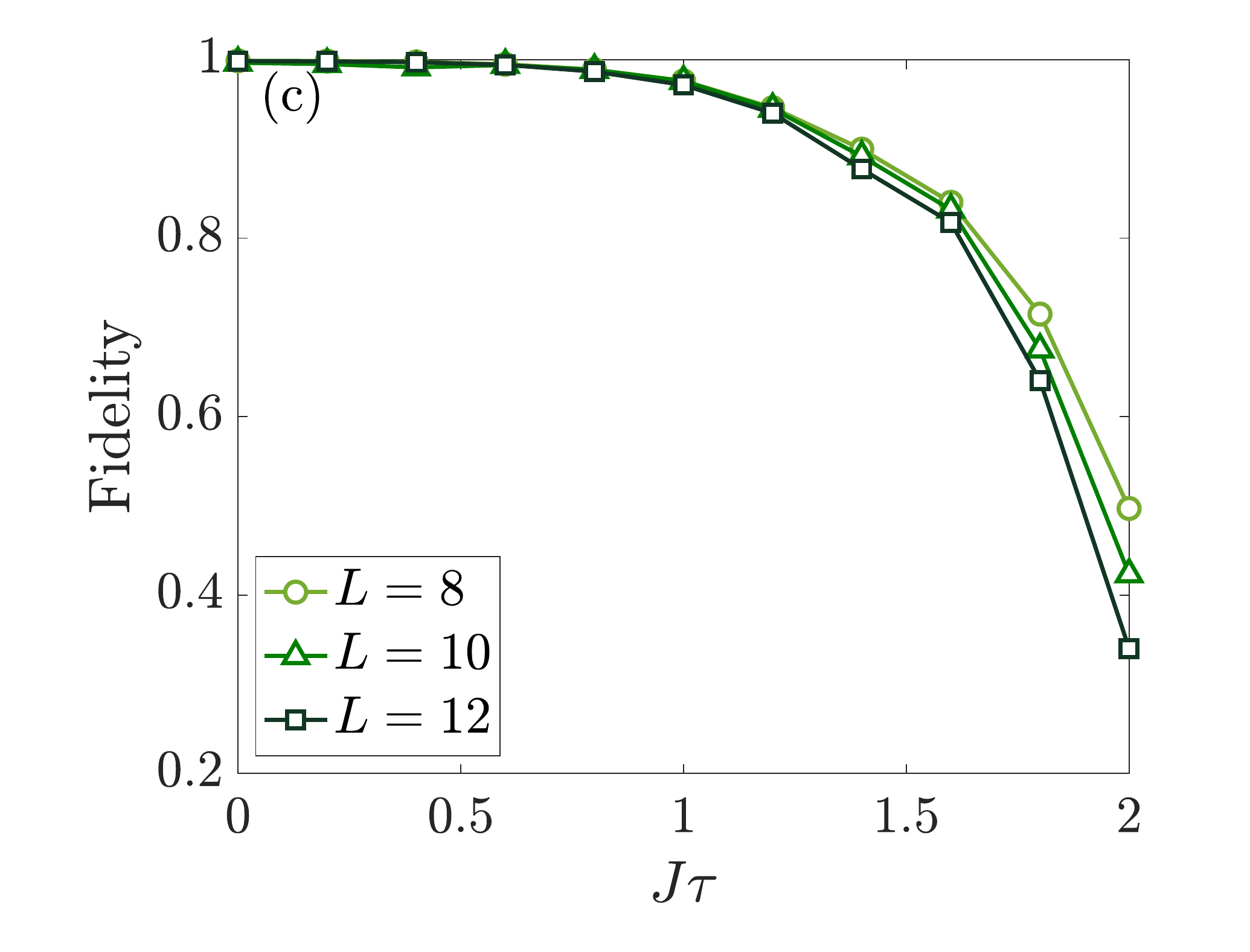}
\caption{\label{Dpre_theory}
$D_{pre}$ expansion of KDM. (a) Norm of the $m^\text{th}$ order term of the quasiconserved observable $D_{pre}$ (normalized by $L2^L$). Different curves stand for $h\tau=J\tau$ from 0 to 2 in steps of 0.2. Darker color represents  smaller $J\tau$. (b) Infidelity $1-\langle D_{pre}(\infty)D_{pre} \rangle$ of infinite-time averaged $D_{pre}$ evaluated up to $m^\text{th}$ order. $L=12$ is used. (c) Fidelity $\langle D_{pre}(\infty)D_{pre} \rangle$ evaluated to $7^\text{th}$ order as a function of $h\tau$ for different system sizes.
}
\end{figure*}

To search for additional conserved observables in KDM we develop a method inspired by the existence of discrete time-translation symmetry-protected phases in prethermal Floquet systems~\cite{Else17}.
Similar results have been obtained  
for the static Hamiltonian $\overline H=hZ+JD_y$ associated with the (zero-order) KDM. 
For this model, it has been shown that the polarization $Z$ is quasiconserved, and does not reach its thermal equilibrium value until a time  exponentially long in $h/J$ \cite{Else17,Abanin17b,Wei19}, even if according to ETH the system should thermalize. 

Since the average Hamiltonian picture breaks down when increasing $\tau$ and we see from Fig.~\ref{brute}(a) that the other observable is conserved for even larger $\tau$, we must go beyond the static case, and work directly in the Floquet system. This kind of system was first studied in \cite{Else17}, where they further focused on the case $h=\pi$ to identify a prethermal Floquet time crystal. 
Here we generalize their analysis to obtain the novel quasiconserved observable for any $h$. 

We transform the Floquet operator by going to a rotated frame as
\begin{equation}\label{eq:pre0}
 e^{S} e^{-i hZ\tau}e^{-i H_1\tau} e^{-S} =  e^{-i h Z\tau} e^{-i\tau (JD+\delta H)},
\end{equation}
and demand $[Z,D]=0$. By appropriately choosing $S,D$, it will be shown that $\delta H$ is exponentially small in $ \min[O(\frac{h}{J}), O(\frac{1}{h\tau})]$~\cite{Peng2019footnote}. Therefore, for small $\tau$ and large enough ratio $h/J\gtrsim0.5$~\cite{Wei19}, the operator $D$ approximately commutes with the Floquet unitary in the rotated frame, making $D_{pre}=e^{-S}De^S$ a prethermal quasiconserved observable in the original frame. 
We emphasize that the right-hand side of Eq.~\ref{eq:pre0} still describes a Floquet system, therefore we derived the quasiconservation without first transforming to a static Hamiltonian.
Note that $Z_{pre}=e^{-S}Ze^S$ is quasiconserved in the same sense as $D_{pre}$. However, whereas $D_{pre}$, is orthogonal to $H_{pre}$ to zeroth order,   $Z_{pre}\approx H_{pre}-D_{pre}$ and it cannot thus be considered an eigen-quasiconserved observable. 

Now we give the details of finding the desired $S,D$. We first write the transformation Eq.~\ref{eq:pre0} in an equivalent form
\begin{equation}\label{eq:pre}
e^{i \epsilon hZ\tau} e^S e^{-i \epsilon hZ\tau} e^{-i \epsilon^2 H_1\tau} e^{-S} = e^{-i\tau (D+\delta H)},
\end{equation}
Here we assume that $J/h$ and $h\tau$ are small parameters whose magnitude are of the same order marked by $\epsilon$, and do perturbation in $\epsilon \ll 1$.
After expanding the operators, $D=\epsilon D_1+\epsilon^2 D_2+\cdots, S = \epsilon S_1+\epsilon^2 S_2+\cdots$, one can collect terms that are of order  $\epsilon^{j}$ on both sides of Eq.~\ref{eq:pre}, and get a series of equations indexed by $j$. In practice we do not calculate exponentials directly but use the Magnus expansion of the left-hand side. The $j$-th order is given by
\begin{equation}\label{eq:pre_j}
    -i\tau D_j = \left[S_{j-1},-ihZ\tau\right] + h_j,
\end{equation}
where $h_j$ only contains $-ihZ\tau$, $-iH_1\tau$ and $S_{j'}$ with $j'<j-1$. The first few orders can be written explicitly, 
\begin{eqnarray}
	&h_1=0,\nonumber \\ 
&h_2=-iH_1\tau,\\ 
&h_3=[S_1,h_2]+\frac{ih\tau}2([S_1,[S_1,-Z]]+[Z,[ihZ\tau,S_1]]),\nonumber
\end{eqnarray}
 while higher orders can be found recursively.
Assuming all orders $S_{j'}$ with $j'<j-1$ are known (which is trivially true for $j=2$), we determine $S_{j-1}$ from Eq.~\ref{eq:pre_j} by requiring $\left[S_{j-1},-ihZ\tau\right]$ to cancel the terms in $h_j$ that do not commute with $Z$. Similar to the prethermal Hamiltonian Eq.~\ref{eq:Hpre}, the expansion in $\epsilon$ generally diverges and should be truncated at some order, leading to the exponentially small nonlocal residual $\delta H$, see, e.g. Ref.~\cite{Abanin17,Else17}.

Here we explain in detail how to obtain $S_{j-1}$ from Eq.~\ref{eq:pre_j} by taking advantage of the special structure of the field operator $Z$.
We first decompose $h_j = \sum_{q=0,\pm1,\cdots} h_{jq}$ such that $[Z, h_{jq}] = q h_{jq}$ ($h_{jq}$ are called the $q$-th quantum coherence of $Z$~\cite{Wei18,Munowitz75,Garttner18}). This decomposition is only possible when the dominant part of the Hamiltonian has equally spaced eigenvalues, such as for the collective rotation $H_2^{(K)}=JZ$ in our case. 
Equation~\ref{eq:pre_j} is then satisfied by setting $-i \tau D_j = h_{j0}$ and $S_{j-1} = i\sum_{q\neq 0} h_{jq}/(hq\tau)$.
We note that $S$ is a sufficiently local operator, $r(S_j)=j$, for KDM with nearest-neighbor interaction.

When $\tau$ is small, the $S_j$ operators are dominated by the $(J/h)^{j}$ term.
Therefore, in the $\tau\to0$ limit, the quasiconserved observable found here for the Floquet model reduces to the prethermal quasiconserved observable of the static Hamiltonian $\overline H^{(K)}$~\cite{Wei19,Else17}, where the expansion is a series of $J/h$ and $\delta \tilde{H}\approx \exp(-O(h/J))$.
In this regime, $D_{pre}=- \frac{1}{2}D_z +O((J/h)^2)$, and the expansion  converges for $h/J \gtrsim 0.5$ (up to truncation at exponentially large order) as shown in Ref.~\cite{Wei19} (Note that here we used $h/J=1$).
Instead, for relatively larger $h\tau$, the $S_j$ operators are  dominated by $(h\tau)^j$ and $\delta \tilde{H}\approx \exp(-O(1/h\tau))$, in agreement with the exponentially slow Floquet heating.

We numerically evaluate the convergence properties of $D_{pre}$ in the KDM [Fig.~\ref{Dpre_theory}(a)], using the  metrics  discussed in the previous section, convergence of the order-by-order expansion terms and infinite-time autocorrelation. 
We find that the series converges up to order 7 in the $h\tau$ regime we are interested in. The infinite-time autocorrelation is close to $1$ at small $\tau$, as shown in Fig.~\ref{Dpre_theory}(b) and (c), confirming that the local truncation of $D_{pre}$ (as obtained by the first few orders)  gives rise to quasiconserved observable $\mathcal E_2^{(K)}$. Comparing these results  to the prethermal Hamiltonian shown in Fig.~\ref{Hid_theory}(b) and (c), we find that (i) the normalized autocorrelation of $D_{pre}$ converges to 1 in a larger parameter range ($J\tau\lesssim 1.6$ for $D_{pre}$ and $J\tau\lesssim 1$ for $H_{pre}$), (ii) the autocorrelation shows a significant drop at $J\tau\gtrsim1.8$ for $D_{pre}$ and $J\tau\gtrsim1.2$ for $H_{pre}$, with a steeper  drop  when $L$ is increased from 8 to 12. Both facts suggest that $D_{pre}$ is more robust than $H_{pre}$, in agreement with the experimental results presented in Ref.~\cite{Peng2019}. This provides evidence that it is possible to realize novel Floquet phases beyond the effective Hamiltonian picture.

\section{Toward infinite temperature: experimental and numerical signatures} \label{sec:limit}

\begin{figure*}[thb]
\centering
\includegraphics[width=0.32\textwidth]{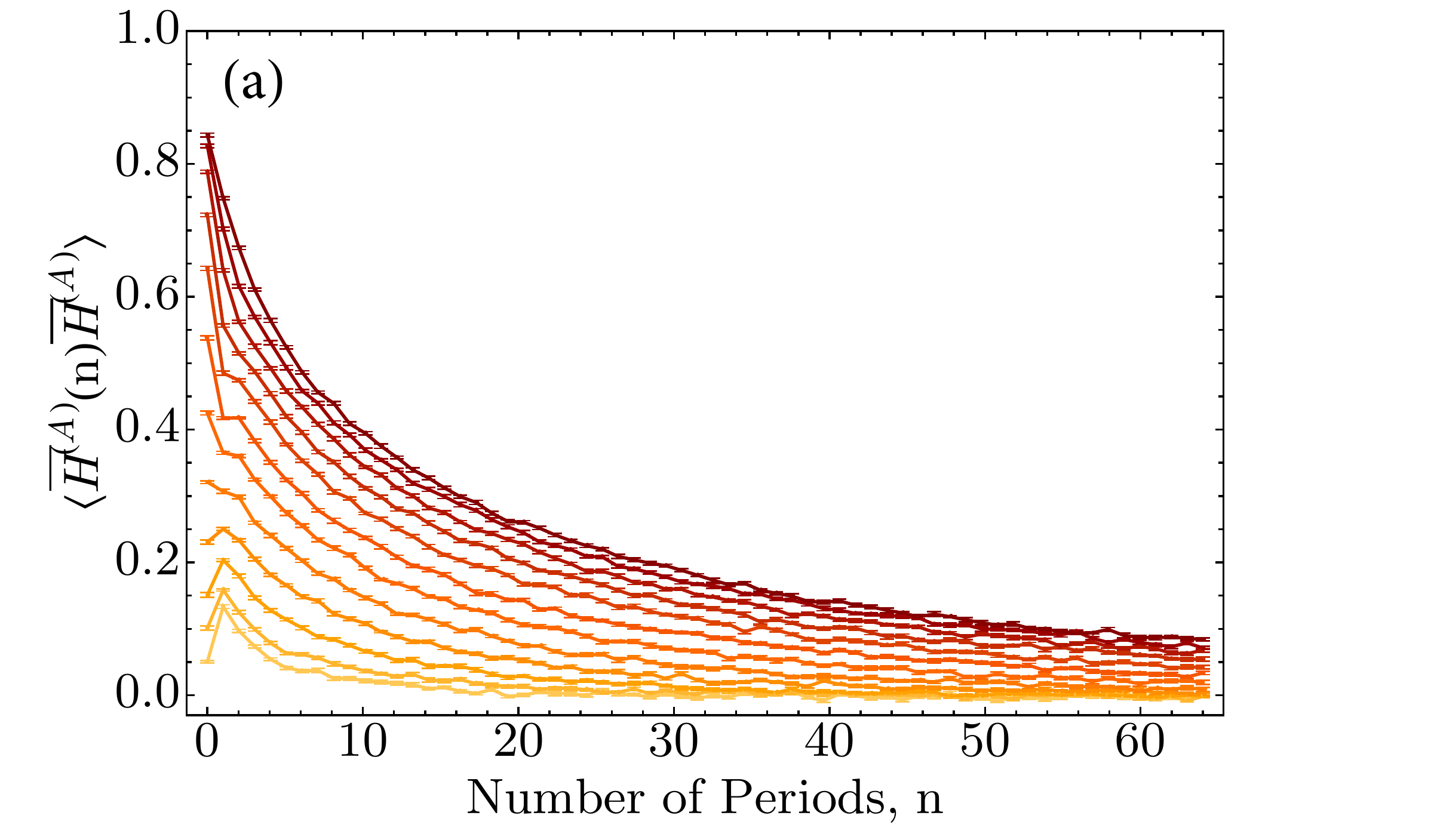}
\includegraphics[width=0.32\textwidth]{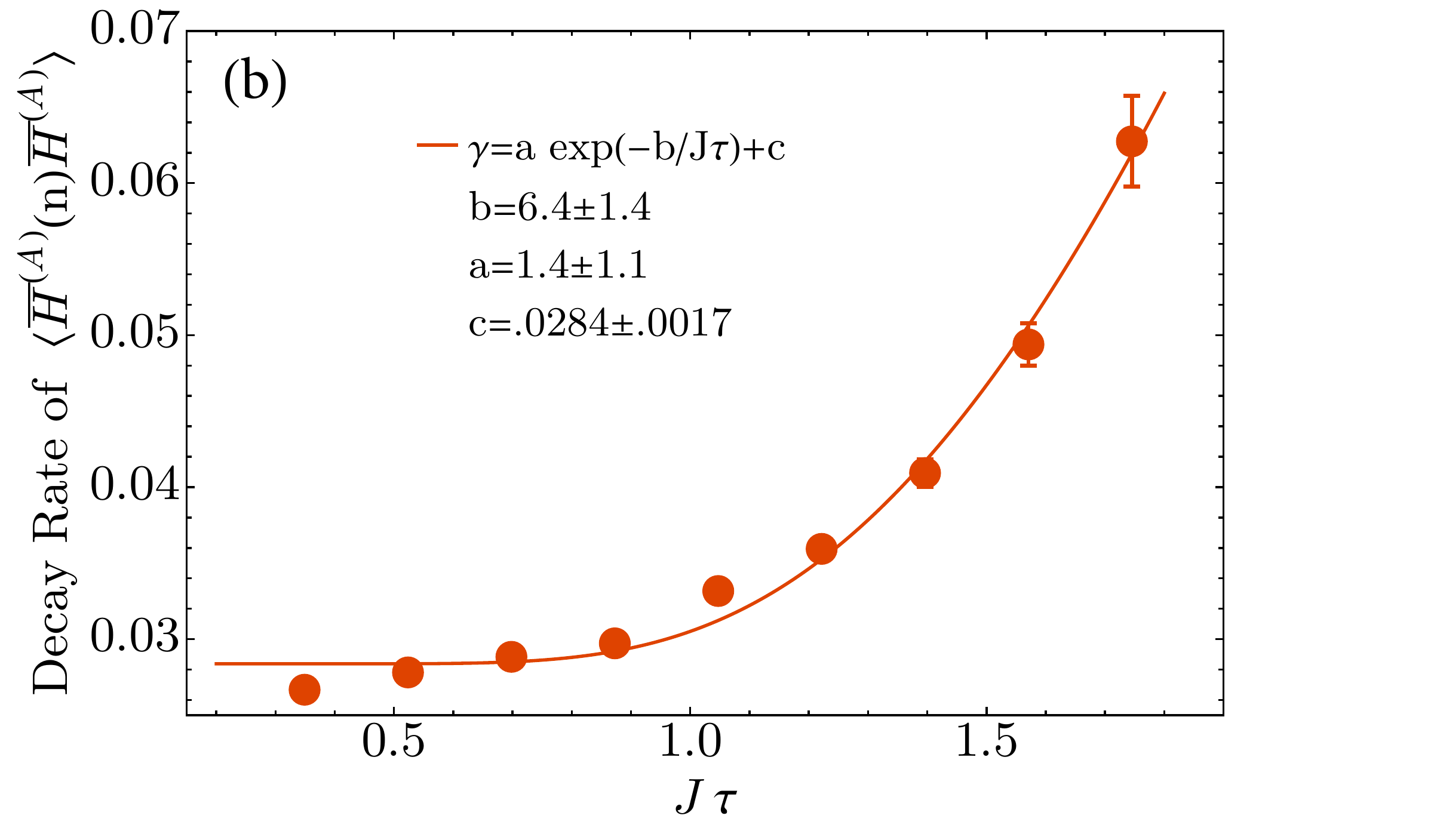}
\includegraphics[width=0.32\textwidth]{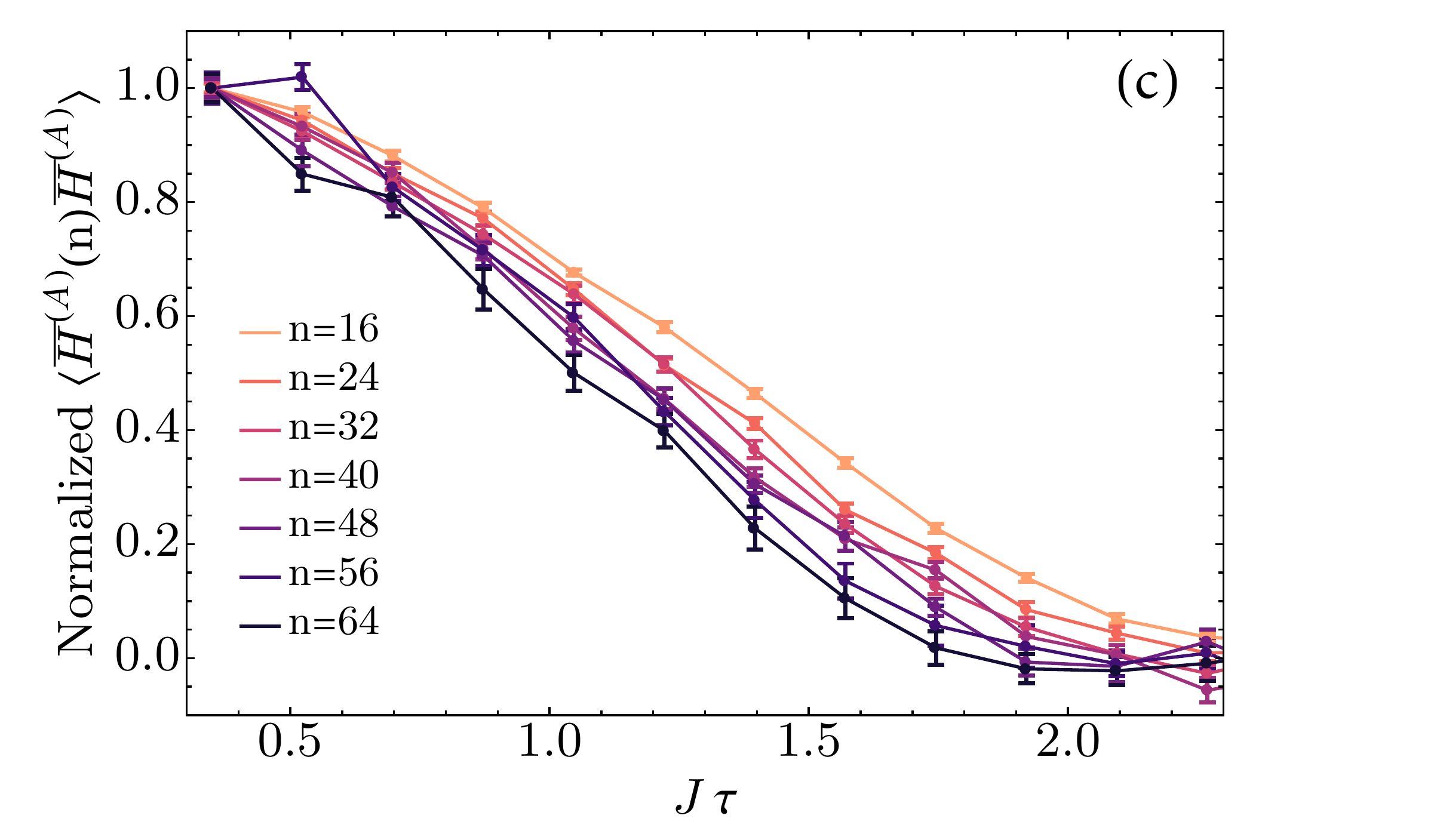}
\caption{\label{fig:ADMexp}
Autocorrelation of the average Hamiltonian for the alternating dipolar model. (a) Autocorrelation as a function of $n$. Different curve stands for $J\tau$ from 0.35 to 2.27 with a step of 0.175. Darker color represents smaller $J\tau$ and lighter color represents larger $J\tau$.
We fit the autocorrelations from $n=20$ to $n=64$ to exponentially decaying function $\exp(-\gamma n)$ and plot the decay rate $\gamma$ in (b). The length of the error bars corresponds to two standard deviation of the fitted decay rate. Solid curve indicates the fit to function $\gamma=a\exp(-b/J\tau)+c$. The fitted coefficients $a,b,c$ are shown in the plot with the 95\% confidence interval.
(c) Autocorrelation versus $J\tau$ for different $n$. Lighter colors represent smaller $n$ and darker colors represent larger $n$. For a given $n$, the autocorrelation is normalized by $\langle \overline H(n)\overline H\rangle$ at $J\tau=0.35$, i.e. the leftmost point is normalized to 1. In (a-b), error bars are determined from the noise in the free induction decay (see
SM~\cite{SOM} for details on the experimental scheme).
}
\end{figure*}

Although it is generally believed that Floquet many-body systems should heat up to infinite temperature, some numerical works~\cite{Heyl19,Sieberer19,Prosen99,DAlessio13} have found signs of non-thermal behavior in various models.
Here we provide evidence of thermalization in the long-time  and thermodynamic limit, using  numerics and experiments in a NMR quantum simulator~\cite{Peng2019,Wei18,Wei19}, respectively.
In simulations, we can access the infinite-time limit  using exact diagonalization, but only for small system sizes. Conversely, the system size in  NMR experiments is large enough to achieve the thermodynamic limit, but the evolution time cannot be too long due to hardware limitation.  Still, by looking at the dynamics for increasingly longer times (experimentally) and larger system sizes (numerically), we can extract insight on the final fate of the Floquet systems.

The experimental system is a single crystal of fluorapatite (FAp)~\cite{VanderLugt64}. We study the dynamics of $^{19}$F spin-$1/2$ using NMR techniques. Although the sample is 3D, $^{19}$F form quasi-1D structure because the interaction within the chain is $\sim$40 times larger than the interaction between different chains~\cite{Cappellaro07l, Zhang09, Ramanathan11}. Average chain length is estimated to be $>50$ and the coherence time of the $^{19}F$ spins is $T_1\approx0.8s$. The sample is placed in 7~T magnetic field where the Zeeman interaction dominates,  thus reducing the $^{19}F$ spins interaction to the secular dipolar Hamiltonian $H=J_0D_z$ with $J_0=-29.7$~krad/s (we define $z$ as the magnetic field direction). While the corresponding 1D, nearest-neighbor  XXZ Hamiltonian is integrable~\cite{Alcaraz87, Sklyanin88, Wang16a},  the experimental $1/r^3$   Hamiltonian can lead to diffusive~\cite{Sodickson95,Zhang98l} and chaotic behavior~\cite{Jyoti17x} in 3D. In the presence of a transverse field,  the system is known to show a quantum phase transition~\cite{Isidori11}. We use 16 RF pulses~\cite{Wei18,Wei19,Peng2019,Sanchez20} to engineer the natural Hamiltonian into $H_1^{(A)}=JD_y$ and $H_2^{(A)}=JD_x$ with tunable $J$. This enables varying the Floquet steps by tuning $J$, while keeping $\tau$ fixed. Then, experimental imperfections such as decoherence and pulse errors remain the same, and we can faithfully quantify the Floquet heating rate. The initial state is a high-temperature thermal state with small thermal polarization in the magnetic field direction $\rho(0)\approx(\mathbb{1}-\epsilon Z)/2^L$ with $\epsilon\approx 10^{-5}$, and the observable is the collective magnetization along x-axis $\mO=X$. As the identity part does not change under unitary evolution and does not contribute to signal, it is convenient to consider only the deviation from the identity $\delta\rho(0)=Z$, which can be rotated to a desired observable $\mO'$. Therefore, the NMR signal is equivalent to an infinite-temperature correlation $\mathrm{Tr}[\delta\rho(t)X]\to \langle\mO'(t)\mO\rangle_{\beta=0}$. 

We experimentally study the heating rates of the quasiconserved observables and their scaling with Floquet period, to reveal the prethermal phase and investigate the eventual heating to infinite temperature.
In Fig.~(\ref{fig:ADMexp}) we show results for ADM (the two quasiconserved observable in KDM show similar behavior as reported elsewhere~\cite{Peng2019}.) 
To study the autocorrelation of $H_{pre}=\overline{H}+O(\tau)$ in ADM, we measure the average Hamiltonian $\overline H^{(A)}\!=\!JD_y\!+\!JD_x\!=\!-JD_z$, since the higher order terms in Eq.~\ref{eq:Hpre} are not accessible. 
We use the Jeener-Broekaert pulse pair~\cite{Jeener67} to evolve the initial state $\delta\rho$ and experimental observable $X$ into $D_z\propto\overline{H}^{(A)}$.
Because of the difference  $H_{pre}-\overline H$, we still expect an initial transient, over a time $\sim \|H_{pre}\|^{-1}$, where the average Hamiltonian thermalizes to the prethermal Hamiltonian. 
When more Floquet periods are applied, the autocorrelation of $D_z$ slowly decays from its prethermal value.

\begin{figure*}[bht] 
\includegraphics[width=0.30\textwidth]{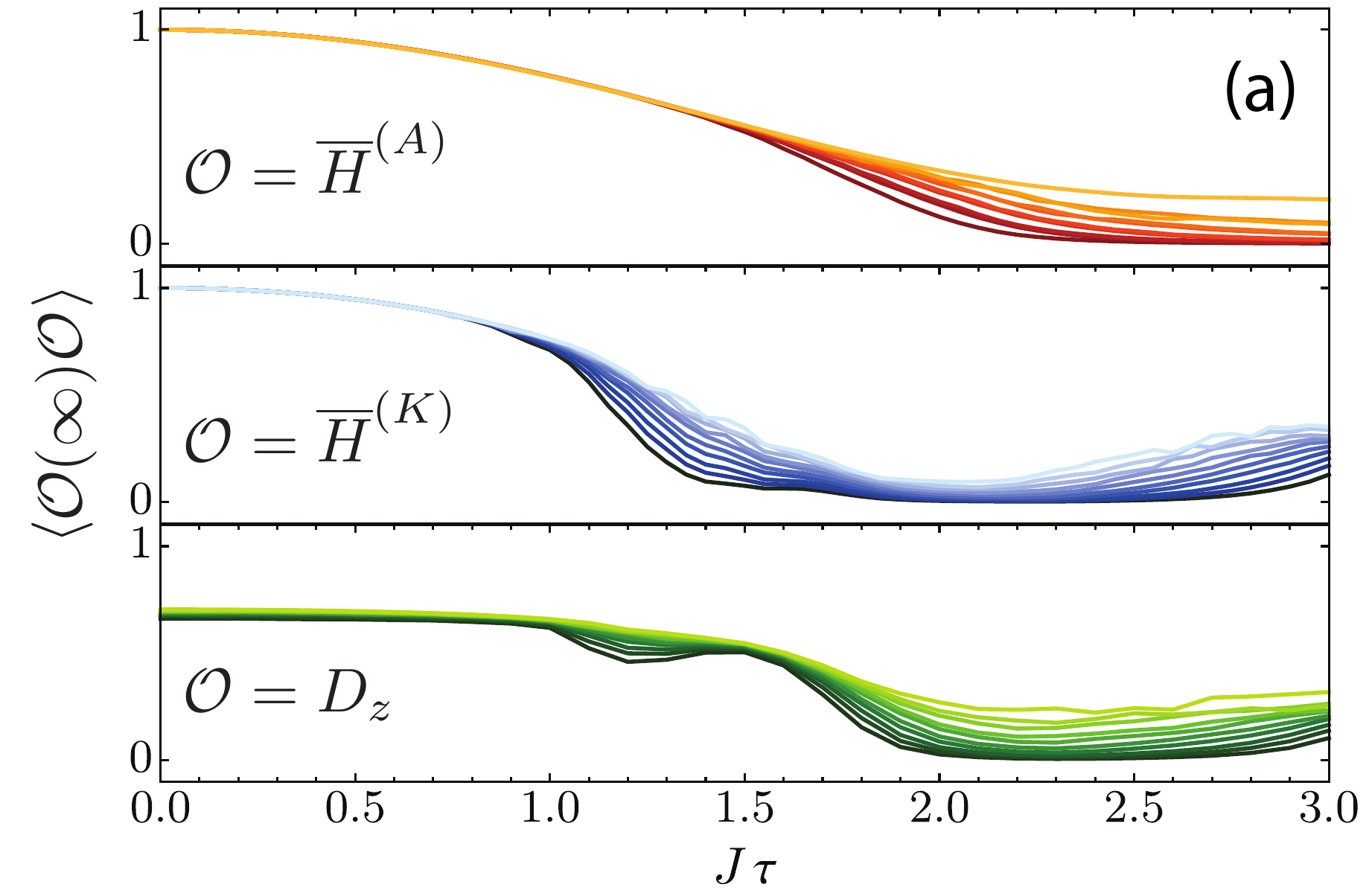}
\includegraphics[width=0.34\textwidth]{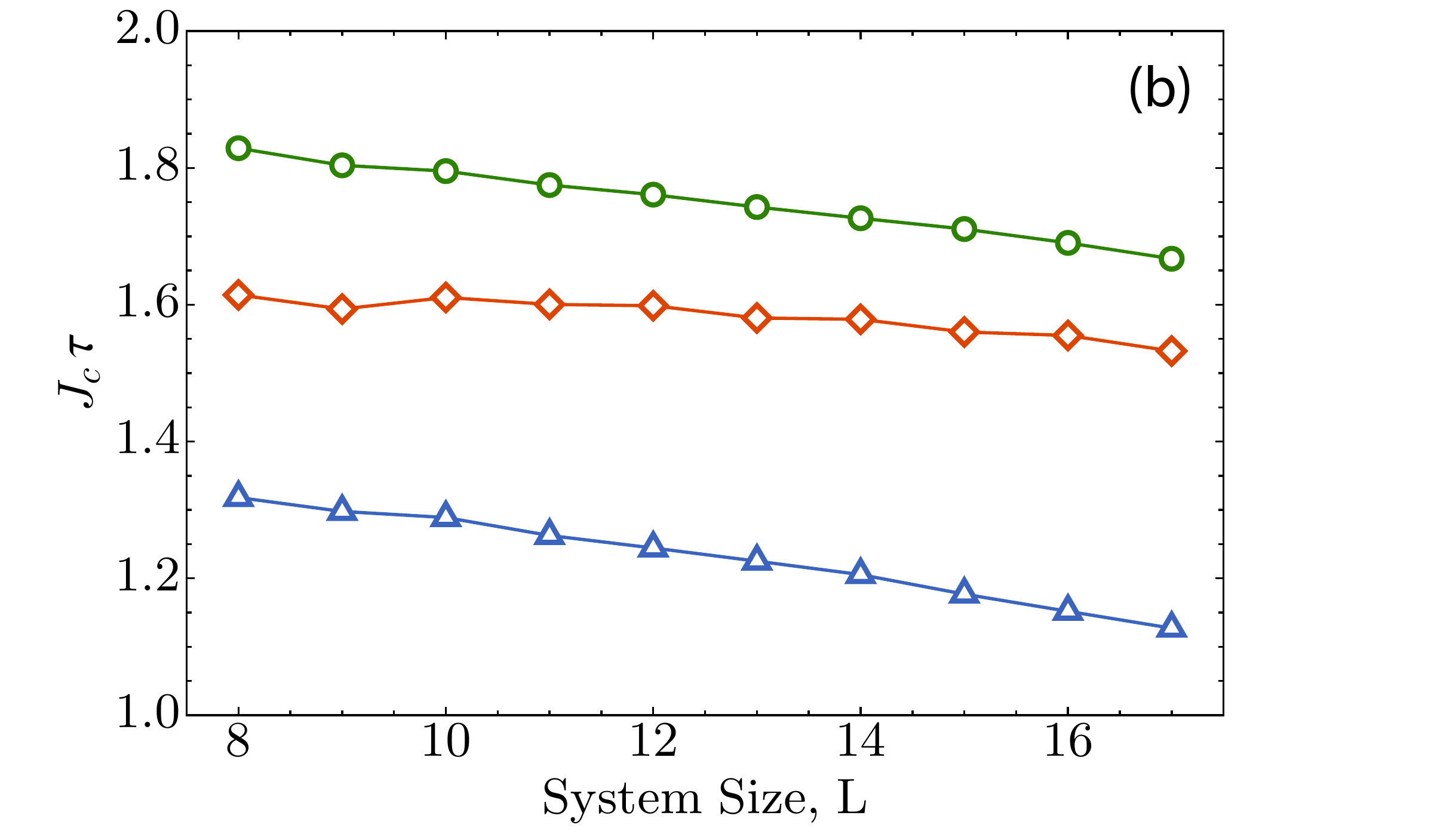}
\includegraphics[width=0.34\textwidth]{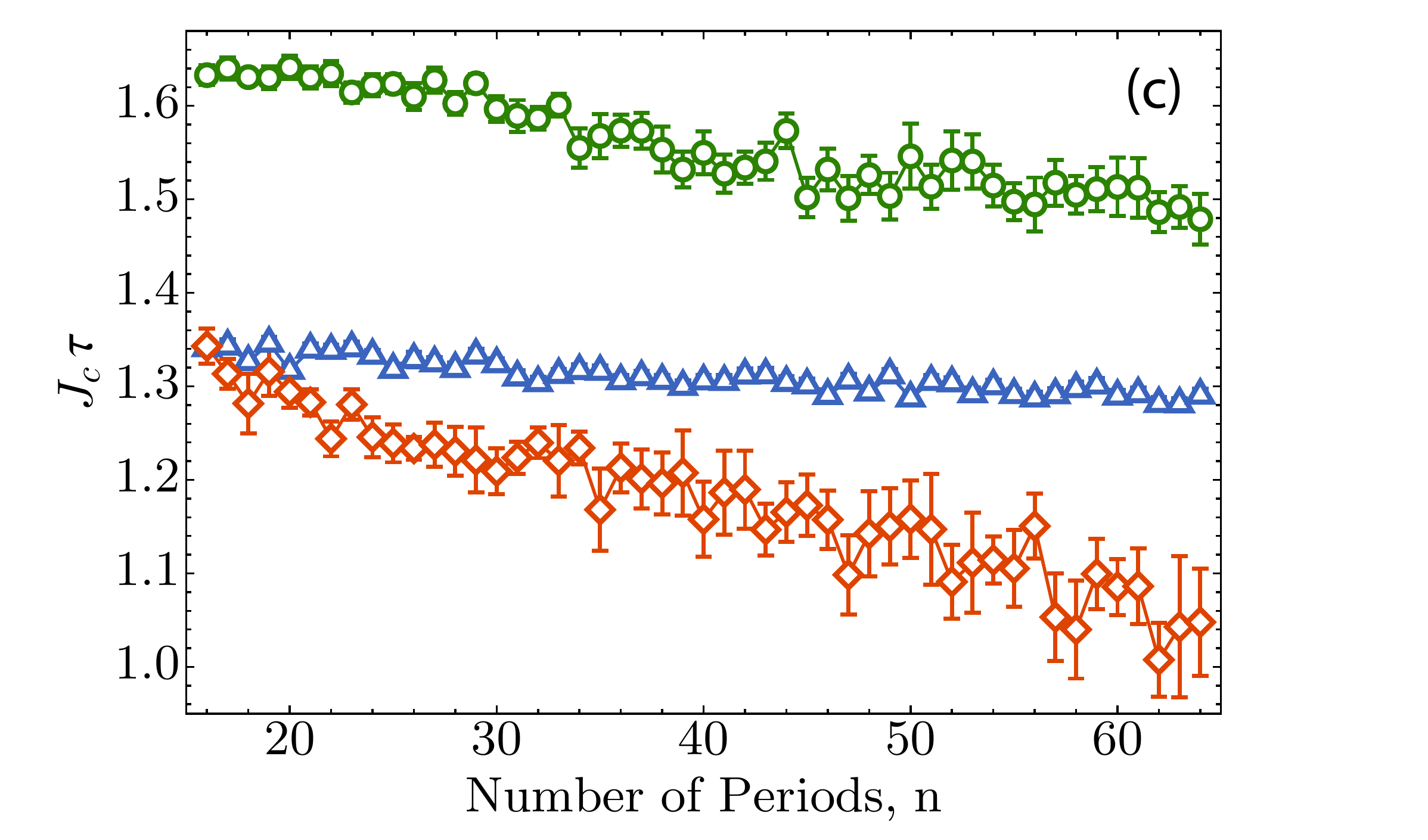}
\caption{\label{fig:tscaling}
Scaling of the critical Trotter step for  KDM  ($\overline H^{(K)}$, blue and $D_z$, green) and  ADM  ($\overline H^{(A)}$, red).
(a) Simulated autocorrelations as a function of $J\tau$ for $L=8,9,\cdots,17$ using exact diagonalization. Darker colors represent larger $L$.
(b) $J_c\tau$ at which the numerical autocorrelation ($L=17$) drops to half of the value under infinitely fast driving ($J\tau\to 0$).
(c) $J_c\tau$ at which the experimentally measured autocorrelation drops to half of the value under the fastest driving ($J\tau=0.35$).
Error bars are determined from the noise in the free induction decay (SM~\cite{SOM}).
}
\end{figure*}

The decay rate in the prethermalization regime is shown in Fig.~\ref{fig:ADMexp}(b), and can be fitted to an exponential function in $1/(J\tau)$ on top of a constant background decay (which is due to experimental imperfections, see SM~\cite{SOM} for more details.)  
By normalizing the data to the data collected under the fastest drive ($J\tau=0.35$), the background decay is cancelled, and the resulting dynamics only arises from the  coherent evolution, as shown in Fig.~\ref{fig:ADMexp}(c). 
For given $n$, the normalized correlation decreases when increasing $J\tau$, because $H_{pre}=\overline{H}+O(J\tau)$ thus $\overline{H}$ that we measure has less overlap with the true quasiconserved observable $H_{pre}$ for larger $J\tau$. The overall drop of the curves when increasing $n$ is instead an indicator of Floquet heating.

To better quantify the final thermalization process, we define a critical value $J_c$ such that when $J\tau>J_c\tau$ the system is thermalized, at a given number $n$  of periods in the thermodynamic limit,  or for a system size $L$ at infinite time.
Studying the scaling of $J_c$ as a function of $n$ (experimentally) and $L$ (numerically) provides hints on the long-time, thermodynamic limits.

We numerically obtain  the autocorrelations $\langle \mathcal O(\infty) \mathcal O\rangle$ as  a function of $J\tau$, using exact diagonalization. In Fig.~\ref{fig:tscaling}(a) we show  simulation results for $\mathcal O=\overline H^{(K)}, D_z$ for KDM and $\mathcal O=\overline H^{(A)}$ for ADM. (Here we explicitly consider the exact dipolar interaction instead of truncating to nearest neighbors.) 
Note that both observables in KDM show a non-monotonic behavior. They appear to be quasiconserved until $J\tau=1$; the decrease in overlap is however interrupted by a revival at  $J\tau=1.6$. This is because $\overline{H}^{(K)}$ and $D_z$ are approximation of $H_{pre}$ and $D_{pre}$ to leading order. Thus $\overline{H}^{(K)}$ ($D_z$) still has a small overlap with $D_{pre}$ ($H_{pre}$), giving rise to a second plateau at $J\tau\approx 1.6$ ($J\tau\approx 1$). 
The experimentally measured autocorrelations of quasiconserved observables in KDM can be find in~\cite{Peng2019}.
For both experiments and simulations we then find $J_c\tau$ from the point where the curves drop below a threshold value of 0.5 (any other reasonable choice would not qualitatively change the results). 
We linearly interpolate between data points to get $J_c\tau$
for every quasiconserved observable and plot the $J_c\tau$ in Fig.~\ref{fig:tscaling}(b) and (c).
The decrease of numerically calculated $J_c\tau$ with $L$ in Fig.~\ref{fig:tscaling}(b) indicates that even the correlations of quasiconserved observables decay to zero as the system thermalizes to infinite temperature, suggesting this non-thermalizing behavior should not persist to the thermodynamic limit. Similar result is also observed from experimentally measured $J_c\tau$ as shown in Fig.~\ref{fig:tscaling}(c)~\footnote{We note that discrepancies in the value of $J_c\tau$ and order of curves  in Fig.~\ref{fig:tscaling}(a) and (c) are to be expected, because although $J_c\tau$ approaches zero when $L\to\infty$ and $n\to \infty$, the convergence speed depends on the path to that limit.}.
Note that although $J_c\tau$ for $\langle \overline H^{(K)}(n)\overline H^{(K)}\rangle$ shows only a moderate dependence on $n$ [Fig.~\ref{fig:tscaling}(c)], its decay is still larger than experimental uncertainties.

\section{Conclusion}\label{sec:conclusion}
As Floquet driving is a promising avenue for quantum simulation, it is crucial to evaluate its robustness, the existence of a long-lived prethermal phase, and the eventual thermalization to infinite temperature. Investigating Floquet heating, which breaks the prethermal regime, is particularly challenging, not only because of inherent limitations in numerical and experimental studies, but also because of the challenge to properly identifying all quasiconserved observables in the complex, many-body driven dynamics.  

Here we tackle both these issues by combining analytical, numerical and experimental tools. First, we provide a systematic strategy to find local, eigen-quasiconserved observables in the prethermal regime using infinite-temperature correlations. 
By systematically searching over local operators, we find that counter-intuitive quasiconserved observables might emerge, as we identify two eigen-quasiconserved observables: the first, not surprisingly is associate with energy, $H_{pre}$, under sufficient fast drive; in addition, we find another quasiconserved observable, $D_{pre}$, for the KDM in the presence of a large driving field.

We then use numerical and experimental evidence to obtain insight into the inaccessible thermodynamic limit and long-time regime, to show that autocorrelations of quasiconserved observables indeed decrease toward zero due to Floquet heating, suggesting the Floquet system approaches the infinite temperature state.

Our results not only provide a metric to study thermalization in driven quantum systems, but also open intriguing perspectives into the existence of   quasiconserved observables other than the energy. It is an open question when they emerge and how  they interact with each other. A better understanding of quasiconserved observables would benefit understanding of heating in closed driven systems, and designing robust protocol to slow down thermalization.

\begin{acknowledgments}
Authors would like to thank H. Zhou, W.-J Zhang and Z. Li for discussion. This work was supported in part by the National Science Foundation under Grants No. PHY1734011, No. PHY1915218, and No.  OIA-1921199.
\end{acknowledgments}

\bibliography{Biblio}

\begin{thebibliography}{92}
\bibitem [{Pen()}]{Peng2019footnote}%
  \BibitemOpen
  \href@noop {} {}\bibinfo {howpublished} {Note that the matrix representation
  of $Z'$ ($\tilde{Z}$) in the frame transformed by $e^{S'}$ ($e^S$) is the
  same as that of $Z$ in the original frame, so they are different physical
  quantities.}\BibitemShut {Stop}%
\end{thebibliography}

\section*{Supplemental Material}
\section{Experimental background decay rate as a function of $J\tau$} \label{app:background}
In the main text we measured the Floquet heating for a periodic, Hamiltonian switching scheme. While it would be easy to change the period by increasing the time between switches, this would lead to experiments performed with different total times or a different number of control operations. In turns, this can introduce variable amount of decoherence and relaxation effects, and of control errors. Instead, we kept the time for one Floquet period constant and used Hamiltonian engineering to vary the Hamiltonian strength in order to vary the Floquet driving frequency. 

\begin{figure}[t]
\centering
\includegraphics[width=0.98\columnwidth]{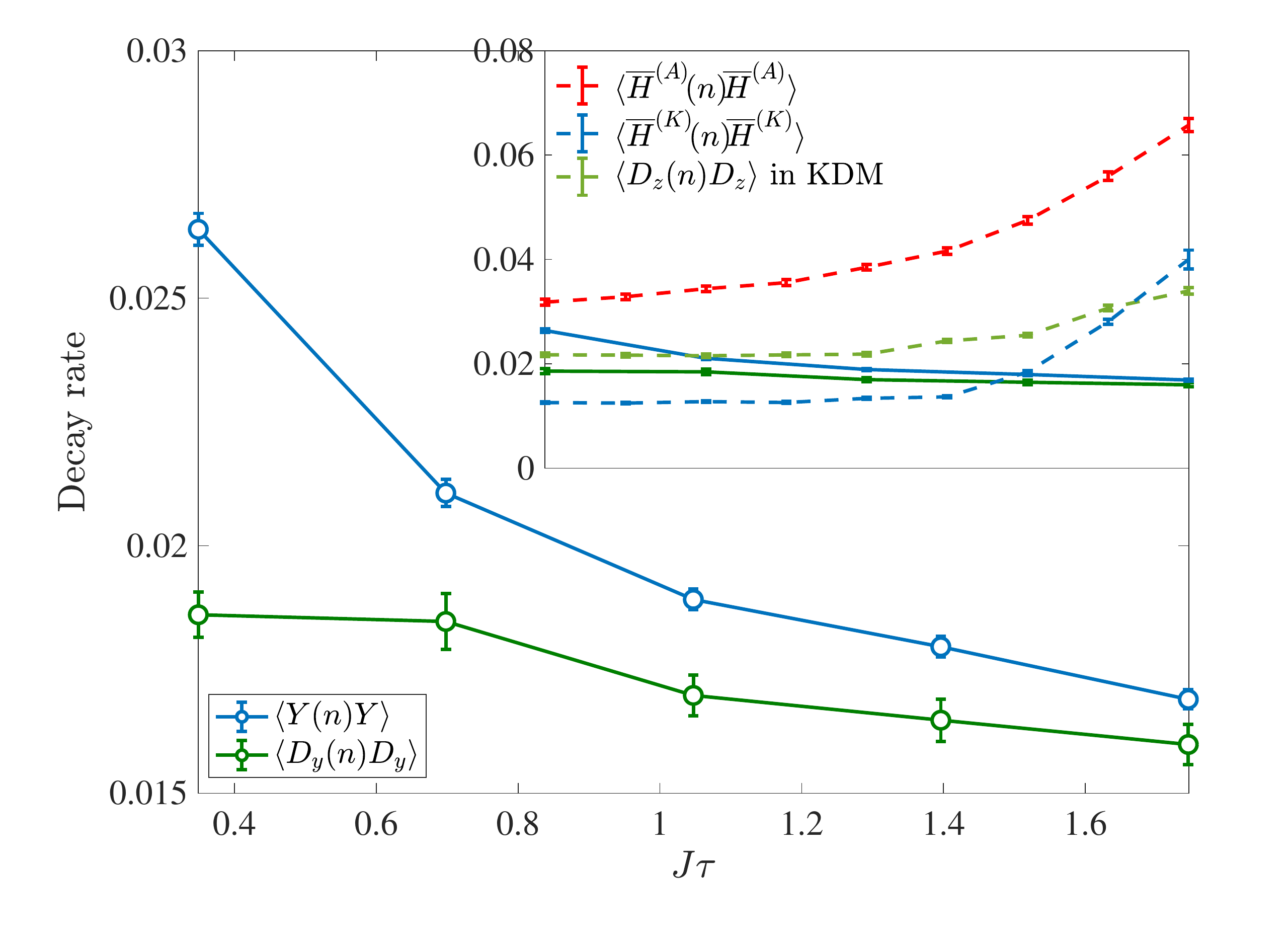}
\caption{\label{fig:bgDR}
Decay rate of $\langle Y(n)Y\rangle$ (blue) and $\langle D_y(n)D_y\rangle$ (green) under engineered dipolar Hamiltonian $JD_y$ as a function of $J\tau$. {The range of $J\tau$ studied was obtained by varying the scaling $u$ (see SM~\cite{SOM}) from 0.098 to 0.646, while keeping fixed $\tau=120\mu s$. In the inset, we compare the background decay rates with the Floquet decay rates (dashed lines).}
}
\end{figure}

One of the assumptions in our work is that the background decay rate does not change much with driving frequency (compared to the change in Floquet heating rate). In this section, we provide experimental evidence for this assertion. 
When changing driving frequency, we are changing (i) the effective strength $J$ of the engineered dipolar interaction $JD_y$ and (ii) the kicking angle in the kicked dipolar model by a phase shift (see \ref{app:Ham}). 
As phase shift angles are usually very accurately implemented in NMR experiments, we focus on the engineered dipolar interaction, which is obtained by Floquet engineering itself, as explained in \ref{app:Ham}.
To quantify how good is the engineered $JD_y$, we measure $\langle Y(n)Y\rangle$ and $\langle D_y(n)D_y\rangle$ under the engineered Hamiltonian $JD_y$, without kicking field nor direction alternation, as shown in Fig.~\ref{fig:bgDR}. 

Note that the maximum difference between the decay rate of $\langle D_y(n)D_y\rangle$ over the range of $J\tau$ considered is $\sim 0.003$, much smaller than the Floquet heating rate in the main text. A quantitative analysis is challenging because the specific form of error terms is unknown, and $JD_y$ is an interacting Hamiltonian thus error accumulation is intractable. 
Here we use some simple arguments to argue that variations in the background decay with $J\tau$ have little to no influence on our results. First, we note that while in the main text we are interested in the decay of the autocorrelation of $H_{pre}$ and $D_{pre}$, here with $H=JDy$ we can only discuss the decay of $D_y$ and $Y$, since other observables that are not conserved display very fast decay which is not informative.
For example, in the main text we measure $D_z$, which thermalizes even under the ideal $D_y$ and thus we cannot distinguish thermalization from decay due to experimental imperfections in the engineered dipolar Hamiltonian $D_y$. Still, as $D_z$ and $D_y$ overlap, if the background decay of $D_z$ had a significant change with $J\tau$, it would be reflected in $D_y$, which is not observed. Therefore, we expect the change in the background decay rate for $\langle D_z(n) D_z\rangle$ to be small as well.
Here we can only probe the background decay rate of $Y$, while in the main text we are interested in the longitudinal magnetization, $Z$, that appears in $\langle \overline H^{(K)}(n)\overline H^{(K)}\rangle$ [see Fig.~\ref{fig:tscaling}(c)]. The transverse magnetization decay rate is, however, a upper bound for $Z$, since in NMR experiments $Z$ is usually more robust against errors than $Y$ due to the large magnetic field in z-axis that suppresses decoherence and experimental errors that do not conserve the total Zeeman energy (we note that we typically do not explicitly write the Zeeman energy in the Hamiltonians as we work in the rotating frame). 
Even if the variation in the background decay for $Z$ were as large as what observed for $Y$ in these experiments ($\sim0.009$), it would still be still small compared with Floquet (see inset of Fig.~\ref{fig:bgDR}). In addition, in the kicked dipolar model, we can consider $ D_y$ as being subjected to rotations along $Z$ that further cancel out the error terms in the engineered $JD_y$ that do not conserve $Z$. 
As a result, the decay rate of $Y$ due to the engineered $D_y$ is larger, by about a factor of 2, than the baseline decay of $\langle \overline H^{(K)}(n)\overline H^{(K)}\rangle$ in the kicked dipolar model (they are 0.254 and 0.123, respectively, in the fastest driving case $J\tau=0.35$).

\section{Experimental System, Control and Data Analysis}
\subsection{Experimental System}
\label{app:exp}
The system used in the experiment was a single crystal of fluorapatite (FAp). Fluorapatite is a hexagonal mineral with space group \(P6_3/m\), with the \(^{19}\)F spin-1/2 nuclei forming linear chains along the \(c\)-axis. Each fluorine spin in the chain is surrounded by three \(^{31}\)P spin-1/2 nuclei.
We used a natural crystal, from which we cut a sample of approximate dimensions 3 mm$\times$3 mm$\times$2 mm.
The sample is placed at room temperature inside an NMR superconducting magnet producing a uniform $B=7$ T field. The total Hamiltonian of the system is given by
\begin{equation}
H_\mathrm{tot}=\omega_F \sum_k S_z^k+\omega_P \sum_\kappa s_z^\kappa+H_{F}+H_P+H_{FP}
\label{eq:Hamtot}	
\end{equation}
The first two terms represent the Zeeman interactions of the F($S$) and P($s$) spins, respectively, with frequencies $\omega_F=\gamma_FB\approx (2\pi)282.37$ MHz and $\omega_P=\gamma_PB=(2\pi)121.51$ MHz, where $\gamma_{F/P}$ are the gyromagnetic ratios. The other three terms represent the natural magnetic dipole-dipole interaction among the spins, given generally by
\begin{equation}
 H_\mathrm{dip}=\sum_{j<k}\frac{\hbar\gamma_j\gamma_k}{|\vec r_{jk}|^3}\left[\vec S_j\cdot\vec S_k-\frac{3\vec S_j\cdot\vec r_{jk}\,\vec S_k\cdot\vec r_{jk}}{|\vec r_{jk}|^2}\right],
\end{equation}
where $\vec r_{ij}$ is the vector between the $ij$ spin pair. Because of the much larger Zeeman interaction, we can truncate the dipolar Hamiltonian to its energy-conserving part (secular Hamiltonian). We then obtain the homonuclear Hamiltonians
\begin{equation}
 \begin{aligned}
 H_F&=\frac{1}{2}\sum_{j<k}J^F_{jk}(2 S_z^j S_z^{k}- S_x^j S_x^{k}- S_y^j S_y^{k}) \\ H_P&=\frac{1}{2}\sum_{\lambda<\kappa}J^P_{\kappa\lambda}(2s_z^\lambda s_z^{\kappa}-s_x^\lambda s_x^{\kappa}-s_y^\lambda s_y^{\kappa})
 \end{aligned}
\end{equation}
and the heteronuclear interaction between the $F$ and $P$ spins,
\begin{equation}
 H_{FP}=\sum_{k,\kappa} J^{FP}_{k,\kappa}S_z^ks_z^\kappa,
\end{equation}
with $J_{jk}=\hbar\gamma_j\gamma_k\frac{1-3\cos(\theta_{jk})^2}{|\vec r_{jk}|^3}$, where $\theta_{jk}$ is the angle between the vector $\vec r_{jk}$ and the magnetic field $z$-axis. The maximum values of the couplings (for the closest spins) are given respectively by $J^F=-32.76$ krad s$^{-1}$, $J^P=1.20$ krad s$^{-1}$ and $J^{FP}=6.12$ krad s$^{-1}$. 
\begin{figure}[b]
\centering 
\includegraphics [width=0.98\linewidth ]{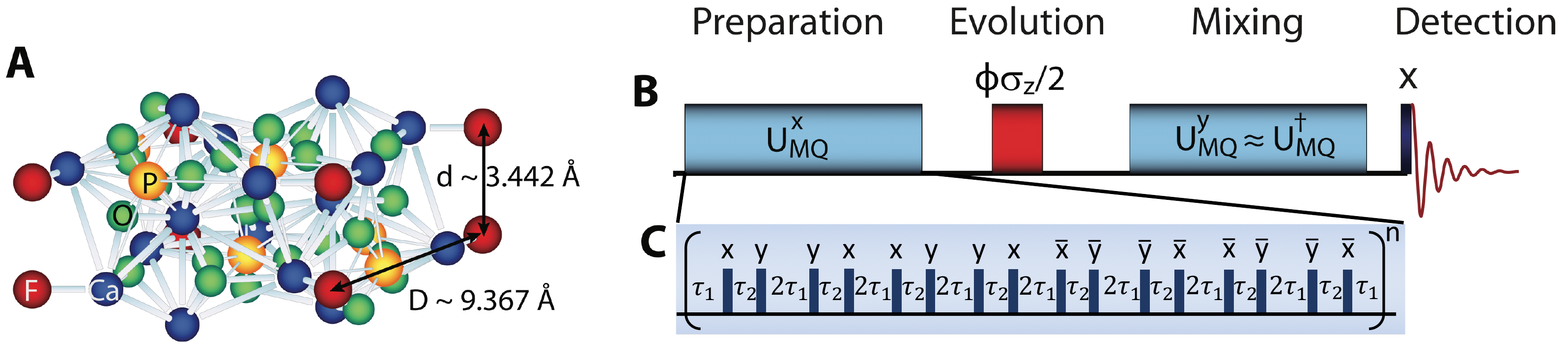}
\caption{\textbf{A} Fluorapatite crystal structure, showing the Fluorine and Phosphorus spins in the unit cell. \textbf{B} NMR scheme for the generation and detection of MQC. In the inset (\textbf{C}) an exemplary pulse sequence for the generation of the $H_\mathrm{dipy}$. Note that thanks to the ability of inverting the sign of the Hamiltonian, the scheme amounts to measuring out-of-time order correlations.
}\label{fig:mqcd}
\end {figure}

The dynamics of this complex many-body system can be mapped to a much simpler, quasi-1D system. First, we note that when the crystal is oriented with its $c$-axis parallel to the external magnetic field
the coupling of fluorine spins to the closest off-chain fluorine spin is $\approx40$ times weaker, while in-chain, next-nearest neighbor couplings are $8$ times weaker. 
 Previous studies on these crystals have indeed observed dynamics consistent with spin chain models, and the system has been proposed as solid-state realizations of quantum wires
~\cite{Cappellaro07l,Cappellaro11,Ramanathan11}. This approximation of the experimental system to a 1D, short-range system, although not perfect has been shown to reliably describe experiments for relevant time-scales~\cite{RufeilFiori09b,Zhang09}. The approximation breaks down at longer times, with a convergence of various effects: long-range in-chain and cross-chain couplings, as well as pulse errors in the sequences used for Hamiltonian engineering. In addition, the system also undergoes spin relaxation, although on a much longer time-scale ($T_1=0.8~$s for our sample). 

\subsection{Error analysis}\label{app:err}
\begin{figure}[h]
\centering
\includegraphics[width=80mm,clip]{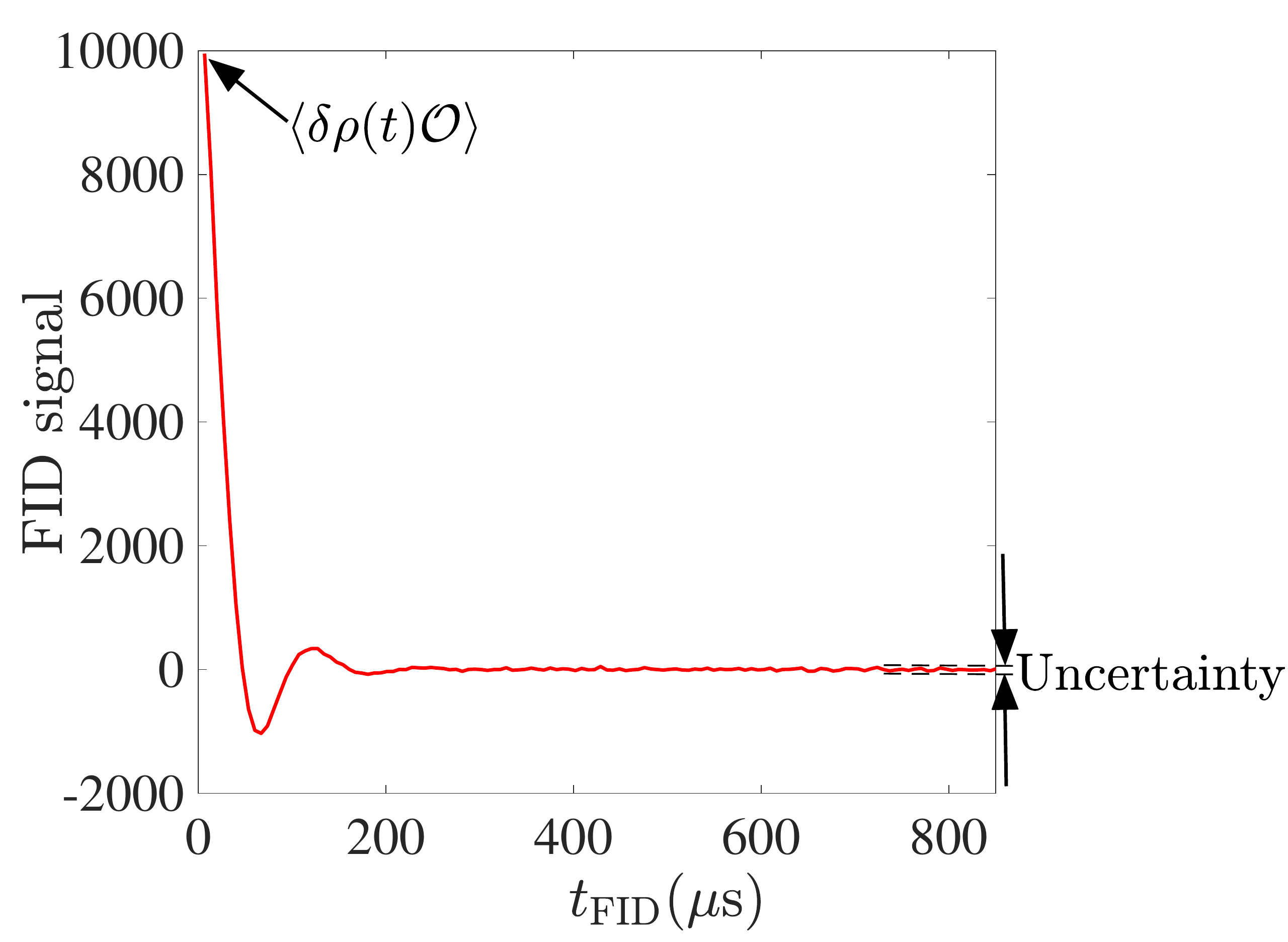}
\caption{\label{fig:FID}
An example of FID. 128 data points are taken in total. The first data point gives $\langle(\delta\rho(t)\mathcal{O}\rangle$ and the standard deviation of the last 20 points gives the uncertainty of $\langle(\delta\rho(t)\mathcal{O}\rangle$.
}
\end{figure}

In experiments, we want to measure the correlation $\langle\delta\rho(t)\mathcal{O}\rangle$, where $\delta\rho(t)= U(t)\delta\rho(0)U(t)$ is the nontrivial part of the density matrix evolved under a pulse-control sequence for a time $t$.
Instead of just performing a single measurement after the sequence, 
we continuously monitor the free evolution of $\delta\rho(t)$ under the natural Hamiltonian $H_\mathrm{dip}$, from $t$ to $t+t_{\textrm{FID}}$.
The measured signal is called in NMR free induction decay (FID) and a typical FID trace is shown in Fig. \ref{fig:FID}). 
This signal trace allows us to extract not only the amplitude of the correlation (from the first data point) but also its uncertainty.
We take the standard deviation of the last 20 data points in the FID as the uncertainty of the $\langle\delta\rho(t)\mathcal{O}\rangle$. This uncertainty is used with linear error propagation to obtain the error bars of all the quantities analyzed in the main text.

\subsection{Hamiltonian Engineering}\label{app:Ham}
In the main text we focused on the Floquet heating (Trotter error) for a periodic alternating scheme, switching between two Hamiltonians. In order to avoid longer times and/or different numbers of control operations when changing the Trotter step (Floquet period), we engineered Hamiltonians of variable strengths. Then, the Hamiltonians themselves are obtained stroboscopically by applying periodic rf pulse trains to the natural dipolar Hamiltonian that describes the system, and are thus themselves Floquet Hamiltonians. Since we only varied the sequences, but not the Floquet period, this step does not contribute to the behavior described in the main text, as we further investigate in ~\ref{app:background}.

We used Average Hamiltonian Theory (AHT~\cite{Haeberlen68}) as the basis for our Hamiltonian engineering method, to design the control sequences and determine the approximation errors. 
The dynamics is induced by the total Hamiltonian \(H=H_\text{dip}+H_\text{rf}\), 
where \(H_\text{dip}=\frac{1}{2}\sum_{j<k}J_{jk}(2 S_z^j S_z^{k}-S_x^j S_x^{k}-S_y^j S_y^{k})+\sum_j h_j S_z^j\) is the system Hamiltonian, 
and \(H_\text{rf}(t)\) is the external Hamiltonian due to the rf-pulses. 
The density matrix \(\rho\) evolves under the total Hamiltonian according to \(\dot\rho=-i[H,\rho]\). 
We study the dynamics into a convenient interaction frame, defined by \(\rho'={U_\text{rf}}^{\dagger}\rho U_\text{rf}\), where \(U_\text{rf}(t)=\mathcal{T}\exp[-i\int_0^t H_\text{rf}(t') dt']\) and \(\mathcal{T}\) is the time ordering operator. 
In this \textit{toggling} frame, \(\rho'\) evolves according to \(\dot{\rho}'=-i[H(t),\rho']\), where \(H(t)={U_\text{rf}}^{\dagger}H_\text{dip} U_\text{rf}\). 
Since \(U_\text{rf}\) is periodic, \(H(t)\) is also periodic with the same period $\tau$, and gives rise to the Floquet Hamiltonian, $H_F$, as as \(U(\tau)=\exp[-i H_F \tau]\). 
Note that if the pulse sequence satisfies the condition \(U_\text{rf}(\tau)=1\), the dynamics of \(\rho\) and \(\rho'\) are identical when the system is viewed stroboscopically, i.e., at integer multiples of \(\tau\), where the toggling frame coincides with the (rotating) lab frame. 

We devised control sequences to engineer a scale-down, rotated version of the dipolar Hamiltonian~\cite{Wei18,Wei19}. We usually look for control sequences that would engineer the desired Hamiltonian up to second order in the Magnus-Floquet expansion.
Then, to engineer the interaction $D_y$, we use a 16-pulse sequence. The basic building block is given by a 4-pulse sequence~\cite{Kaur12,Yen83} originally developed to study MQC.
We denote a generic 4-pulse sequence as \(P(\tau_1,{\bf n}_1,\tau_2,{\bf n}_2,\tau_3,{\bf n}_3,\tau_4,{\bf n}_4,\tau_5)\), where \({\bf n}_j\) represents the direction of the \(j\)-th \(\pi/2\) pulse, and \(\tau_j\)'s the delays interleaving the pulses. In our experiments, the \(\pi/2\) pulses have a width \(t_w\) of typically 1 \(\mu\)s. \(\tau_j\) starts and/or ends at the midpoints of the pulses (see also Fig.~\ref{fig:mqcd}). In this notation, our forward 16-pulse sequence can be expressed as
\begin{widetext}
\begin{gather*}
P(\tau_1,{\bf x},\tau_2,{\bf y},2\tau_1,{\bf y},\tau_2,{\bf x},\tau_1)P(\tau_1,{\bf x},\tau_2,{\bf y},2\tau_1,{\bf y},\tau_2,{\bf x},\tau_1)P(\tau_1,{\bf \overline{x}},\tau_2,{\bf \overline{y}},2\tau_1,{\bf \overline{y}},\tau_2,{\bf \overline{x}},\tau_1)P(\tau_1,{\bf \overline{x}},\tau_2,{\bf \overline{y}},2\tau_1,{\bf \overline{y}},\tau_2,{\bf \overline{x}},\tau_1)
\end{gather*}
and the backward sequence as
\begin{gather*}
P(\tau_3,{\bf y},\tau_3,{\bf x},2\tau_4,{\bf x},\tau_3,{\bf y},\tau_3)P(\tau_3,{\bf y},\tau_3,{\bf x},2\tau_4,{\bf x},\tau_3,{\bf y},\tau_3)P(\tau_3,{\bf \overline{y}},\tau_3,{\bf \overline{x}},2\tau_4,{\bf \overline{x}},\tau_3,{\bf \overline{y}},\tau_3)P(\tau_3,{\bf \overline{y}},\tau_3,{\bf \overline{x}},2\tau_4,{\bf \overline{x}},\tau_3,{\bf \overline{y}},\tau_3)
\end{gather*}
\end{widetext}
where \(\{{\bf \overline{x}},{\bf \overline{y}}\}\equiv \{{\bf -x},{\bf -y}\}\). The delays are given by
\begin{gather*}
\begin{aligned}
\tau_1&=\tau_0(1-u), \quad
\tau_2=\tau_0(1+2u), \\
\tau_3&=\tau_0(1+u), \quad
\tau_4=\tau_0(1-2u),
\end{aligned}
\end{gather*}
where \(\tau_0\) is 5 \(\mu\)s in this paper. The cycle time \(t_c\), defined as the total time of the sequence, is given by \(\tau=24\tau_0\). \(u\) is a dimensionless adjustable parameter, and is restricted such that none of the inter-pulse spacings becomes negative. To the zeroth order Magnus expansion, the above sequence realizes Hamiltonian $uJ_0D_y$ and $uJ_0=J$. 

\end{document}